%% file: Buckling_Main.tex
\documentclass[sn-nature,Numbered]{sn-jnl}

\usepackage{graphicx}%
\usepackage{multirow}%
\usepackage{amsmath,amssymb,amsfonts}%
\usepackage{amsthm}%
\usepackage{mathrsfs}%
\usepackage[title]{appendix}%
\usepackage{xcolor}%
\usepackage{textcomp}%
\usepackage{manyfoot}%
\usepackage{booktabs}%
\usepackage{algorithm}%
\usepackage{algorithmicx}%
\usepackage{algpseudocode}%
\usepackage{listings}%

\usepackage{import}

\begin{document}

\title[Article Title]{A Universal Localized Failure Mechanism for Real Cylindrical Shells}

\author*[1, 2]{\fnm{Nicholas L.} \sur{Cuccia}}\email{cuccia@g.harvard.edu}
\equalcont{These authors contributed equally to this work.}

\author*[]{\fnm{Marec} \sur{Serlin}}
\email{marecserlin@gmail.com}
\equalcont{These authors contributed equally to this work.}

\author[3]{\fnm{Kshitij K.} \sur{Yadav}}

\author[2]{\fnm{Sagy} \sur{Lachmann}}

\author[4]{\fnm{Simos} \sur{Gerasimidis}}

\author*[2]{\fnm{Shmuel M.} \sur{Rubinstein}}\email{shmuel.rubinstein@mail.huji.ac.il}

\affil[1]{\orgdiv{School of Engineering and Applied Sciences (SEAS)}, \orgname{Harvard Universtiy}, \orgaddress{\city{Cambridge}, \postcode{02138}, \state{MA}, \country{USA}}}

\affil[2]{\orgdiv{The Racah Institute of Physics}, \orgname{The Hebrew University of Jerusalem}, \orgaddress{\city{Jerusalem}, \postcode{91904}, \state{Givat Ram}, \country{Israel}}}

\affil[3]{\orgdiv{Department of Civil Engineering}, \orgname{Indian Institute of Technology (BHU)}, \orgaddress{\city{City}, \postcode{221005}, \state{Varanasi}, \country{India}}}

\affil[4]{\orgdiv{Department of Civil and Environmental Engineering}, \orgname{University of Massachusetts}, \orgaddress{\city{Amherst}, \postcode{01003}, \state{MA}, \country{USA}}}

\abstract{\textbf{From cell development to space rockets, the mechanical stability of thin shells is crucial across many industrial and natural processes.  However, predicting shells' failure properties remains an open challenge, owing to their sensitivity to imperfections inevitably present in real structures. Existing predictive methods are informed by classical stability analyses of idealized shells, which associate shell buckling with a spontaneously emerging delocalized buckling mode.  Here, we show through experimental and numerical observations how realistic geometric imperfections instead result in a universal localized buckling mode.  The profile of this mode is robust across shells despite their differing underlying defect structures.  Further contrasting with classical predictions, a real shell's equilibrium surface deformations grow in the shape of this universal localized mode even at stable loads well before failure.  Our findings show that these complex nonlinear dynamics for loads below buckling are described by a 1-D relationship for structures buckling from a saddle-node bifurcation instability.  This breakthrough provides a simplified yet general understanding of buckling mechanisms, leading to improved non-destructive predictive methods.  To showcase these applications, we experimentally implement a lateral-probing technique that predicts the buckling loads of several unmodified shells to within 1\% of the correct value.}}

\keywords{buckling mode, thin cylindrical shells, defect sensitivity, saddle-node bifurcation}

\maketitle
% \section*{Main}

The safe and efficient implementation of mechanical structures in a multitude of engineering contexts hinges on being able to predict when and how they will fail \cite{tsapisOnsetBucklingDrying2005, sacannaLockKeyColloids2010,reisPerspectiveRevivalStructural2015, djellouliShell2024}.
Accurate predictions are particularly crucial for all modern structures constrained to be both lightweight and strong, as the failure of under-engineered designs risks having disastrous repercussions \cite{weingartenvBucklingThinwalledCircular2020}.
Lightweight structures commonly make use of shells---very thin curved plates---because they are a structural element offering an exceptional strength-to-weight ratio.
Despite their ubiquity and importance, the development of reliable methods for predicting failure in shells remains limited by ambiguities in their failure mechanism \cite{thompsonAdvances2015, elishakoffProbabilisticResolutionTwentieth2012, ankalhopeNondestructivePredictionBuckling2022}.
Thus, improving our understanding of how and why shells fail is an essential prerequisite for the efficient design of generic lightweight structures, everything ranging from the ordinary soda can to the extraordinary spacecraft hull.

Classical linear stability analysis provides the conceptual and mathematical framework for describing the tipping point associated with structural failure by relating it to a critical buckling load and buckling mode \cite{thompsonBasic1963, timoshenkoTheory1972, brushBuckling1975, singerBuckling2001}.
A mechanical structure is said to buckle at the smallest load for which a deformation, whose shape defines the buckling mode, can grow without energetic cost.
Traditionally, this theoretical approach assumes that structures are perfect, without any flaws or defects, and predicts a buckling load associated with the spontaneous emergence of a distinctive periodic buckling mode, often resembling a checkerboard pattern \cite{koiterEffect1963, timoshenkoTheory1972, brushBuckling1975, singerBuckling2001, hutchinsonPostbucklingTheory1970}.
Experimentally, however, shells are observed to fail at loads that are highly variable (even within a single manufacturing process) and well below those predicted for their perfect analogs \cite{almrothExperimental1964, tsienTheoryBucklingThin1942, leeGeometric2016}.
Furthermore, high-speed videography ubiquitously reveals that failure initiates via a localized deformation that usually takes on a single diamond-like shape \cite{almrothExperimental1964, singerBuckling2002, huhneNewApproachRobust2005, huhneRobust2008, cucciaHitting2023}.
The theoretically predicted checkerboard pattern only appears later, after the initial localized buckle has triggered non-equilibrium dynamics leading to deformations that encompass the entire shell's surface \cite{almrothExperimental1964, huhneNewApproachRobust2005, singerBuckling2002, cucciaHitting2023}.

Real shells, unlike their idealized counterparts, are not perfect; each individual structure hosts a unique array of defects with a random distribution \cite{hilburgerShell2006, hilburgerBuckling2008, kriegesmannEffectsGeometricLoading2012, kalninsExperimentalNondestructiveTest2015, leeGeometric2016, gerasimidisEstablishingBucklingKnockdowns2018, weingartenvBucklingThinwalledCircular2020}. 
Their buckling loads are particularly sensitive to geometric imperfections, i.e. small variations in radius and thickness, due to shells' overall curvature giving rise to strong non-linear dynamics \cite{donnellEffect1950, koiterEffect1963, horakCylinder2006,weingartenvBucklingThinwalledCircular2020, gerasimidisDentImperfectionsShell2021,royerProbing2022, royerExperimentally2023}.
Even so, analytical investigations of shells hosting imperfections predict global buckling modes \cite{koiterEffect1963}, in contrast to the localized deformations observed in high-speed videos \cite{cucciaHitting2023}.
Present-day efforts to resolve this ambiguity primarily address the emergence of localized deformations in a shell's complex post-buckling dynamics \cite{horakCylinder2006, hilburgerShell2006, hilburgerBuckling2008, kriegesmannEffectsGeometricLoading2012, haynieValidationLowerBoundEstimates2012, castroGeometricImperfectionsLowerbound2014, schultzTest2018, kreilosFully2017, grohRole2019,audolyLocalization2020, huhneNewApproachRobust2005}.
However, recent experimental and numerical studies indicate that the localized post-buckling deformations evolve continuously from localized pre-buckling deformations in shells with realistic imperfections \cite{sunDigitalImageCorrelationaided2022, cucciaHitting2023}.
The inability of the existing conceptual framework to contextualize these observations highlights a fundamental gap in our understanding of the failure mechanism in real shells.

\subsection*{Buckling of Shells with Imparted Localized Imperfection}\label{modified}

We first examine a cylindrical shell whose surface is modified, as described in our previous work \cite{gerasimidisDentImperfectionsShell2021,yadavNondestructive2021,cucciaHitting2023}, to contain a large dimple-shaped geometric imperfection.
Such a modified structure is ideal for the investigation of localization in buckling because the imposed imperfection dominates over the shell's inherent defects, providing a well-defined location for buckling to initiate \cite{gerasimidisDentImperfectionsShell2021, abramianNondestructive2020,yadavNondestructive2021, cucciaHitting2023, leeGeometric2016}.
Moreover, a modified shell allows for careful characterization of surface deformations \textit{prior to} failure because its buckling load can be accurately and non-destructively predicted using the recently developed procedure called \textit{ridge-tracking} \cite{gerasimidisDentImperfectionsShell2021, virotStability2017, abramianNondestructive2020, yadavNondestructive2021, cucciaHitting2023}.

\begin{figure}[h!]
\centering\includegraphics[width=0.9\textwidth]{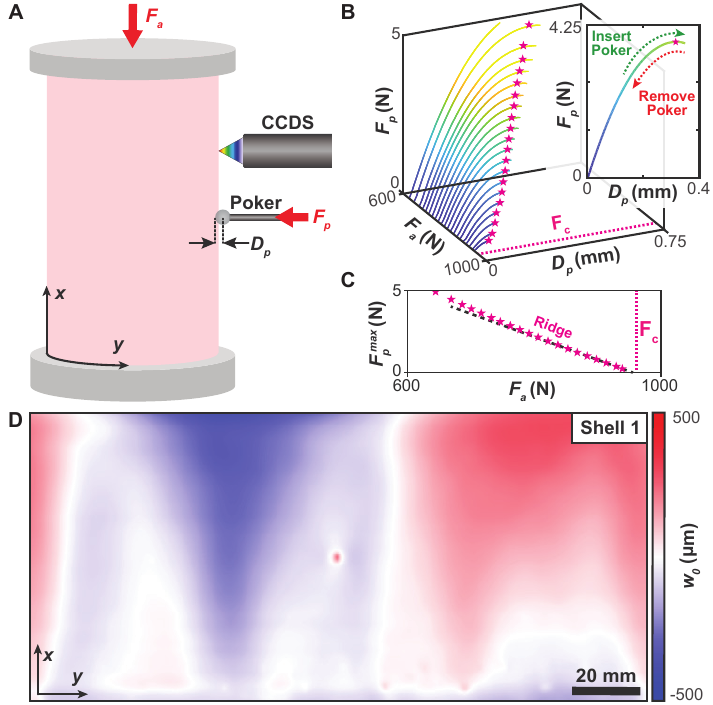}
\caption{\textbf{Lateral probing of shells with geometric imperfections.}
$\textbf{(A)}$ Schematic illustration of a cylindrical shell under an axial load, $F_a$, applied by parallel plates on either end.
Non-contact surface profilometry measurements are performed by rastering a chromatic confocal distance sensor (CCDS) around the shell. The lateral probing response is measured by pushing a poker radially inwards a distance, $D_p$, and measuring the resulting force, $F_p$. $\textbf{(B)}$ The poker force, $F_p$, generated by probing, as a function of its displacement, $D_p$, and the axial load, $F_a$.  During \textit{ridge tracking}, $D_p$ is increased until the first maximum of each probe-displacement curve, $F^{\text{max}}_{p}$, is found (marked with stars). $\textbf{(C)}$ $F^{\text{max}}_{p}$ as a function of the axial load, $F_a$. The buckling load, $F_c$, is predicted from the zero intercept of a linear fit to the load dependence of $F^{\text{max}}_{p}$. $\textbf{(D)}$ The experimentally measured geometric imperfections, $w_0$, of a typical 7.5 oz aluminum soda can with a large imposed localized defect. The imposed defect, visible in the center of the image, has a radius approximately $4\%$ the circumference of the shell and is large in amplitude, with a depth greater than $5$-times the thickness of the shell. Positive values of $w_0$ correspond to deviations pointing radially inwards.}\label{F_1}
\end{figure}

Ridge-tracking is performed by measuring a shell’s response to lateral poking at various axial loads, $F_a$, gradually approaching the buckling load, $F_c$.
At each axial load, the lateral force on the poker, $F_p$, is measured as a function of the inwards radial displacement, $D_p$.
Typically, for axial loads in the range $0.6F_c < F_a < F_c$, the probe's force-response is non-monotonic: the force required to push inwards initially increases to a peak value, $F^{\text{max}}_p$, and then decreases, as shown in the inset of Fig.~\ref{F_1}B.
The relationship between $F_p$, $D_p$, and $F_a$ defines a surface in a three-dimensional phase space, termed the \textit{stability landscape} \cite{virotStability2017}, as shown in Fig.~\ref{F_1}B.
This landscape exhibits a ridge, $F^{\text{max}}_p(D_p,F_a)$, whose height decreases as $F_a$ increases.
When probing at the imparted imperfection, the ridge is empirically observed to linearly trend to zero as $F_a$ approaches $F_c$ \cite{virotStability2017, abramianNondestructive2020, yadavImperfectionInsensitiveThin2020, cucciaHitting2023, leeGeometric2016}.
Thus, using linear extrapolation, \textit{ridge-tracking} enables the non-destructive prediction of the shell's buckling load accurately to within 1\% \cite{leeGeometric2016, marthelotBuckling2017, abramianNondestructive2020,yadavNondestructive2021, abbasiProbing2021, ankalhopeNondestructivePredictionBuckling2022, sunDigitalImageCorrelationaided2022, cucciaHitting2023}, as shown in Fig.~\ref{F_1}C.
Lateral probing was originally devised to measure a shell's energy barrier to spontaneous failure which, in perfect shells, is associated with a non-linear localized \textit{mountain pass mode} \cite{thompsonQuantified2014, thompsonAdvances2015, thompsonShockSensitivity2016, hutchinsonNonlinear2017, hutchinsonImperfections2018, virotStability2017,abramianNondestructive2020, horakCylinder2006, kreilosFully2017, grohRole2019}.
However, this theory, formulated for defect-free shells, does not account for the occasional success of \textit{ridge-tracking} on imperfect shells, nor does it provide guidance on accurately identifying the optimal location for probing.
Indeed, recent studies \cite{cucciaHitting2023,abramianNondestructive2020} show that \textit{ridge-tracking} accurately predicts the buckling load only when probing is performed at the location where buckling initiates. 

\begin{figure}[h!]
\centering\includegraphics[width=0.9\textwidth]{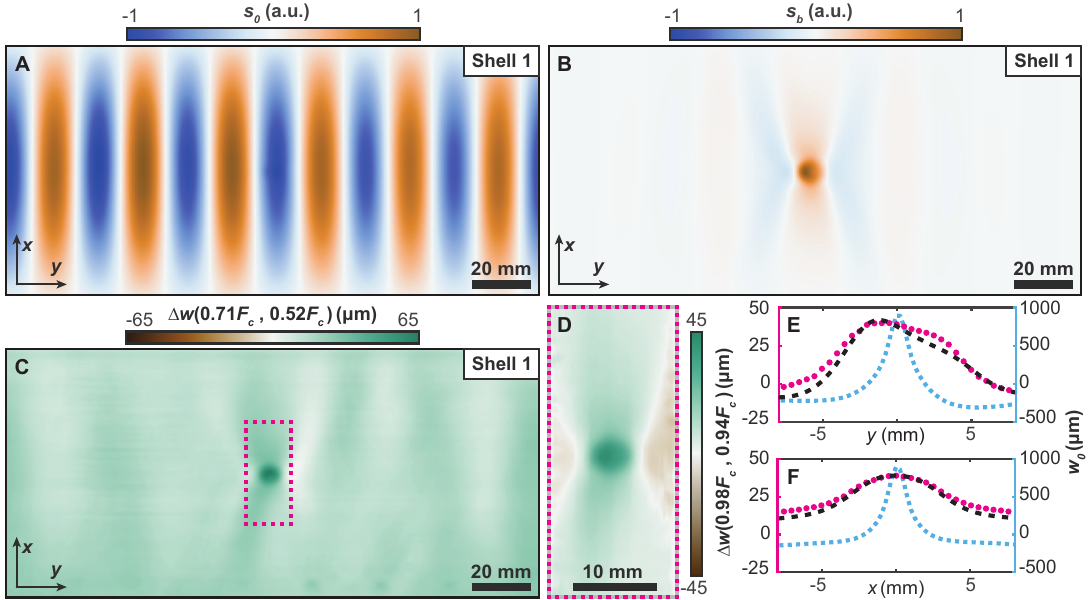}
\caption{\textbf{Calculated eigenmodes and measured deformations of a shell with a single large localized imperfection.}
Eigenmode with smallest eigenvalue when $F_a = 0$ $\textbf{(A)}$ and $F_a = F_c$ $\textbf{(B)}$ calculated numerically using the von Karman Donnell equations for the shell presented in Fig.~\ref{F_1}D.
In the simulation, the eigenmode shown in (B) has an eigenvalue that approaches zero, indicating that it is the shell's predicted buckling mode (supplementary text).
$\textbf{(C)}$ Radial deformations, $\Delta w$, measured when the axial load is increased from $F_a = 0.52 F_c$ to $F_a = 0.71 F_c$. 
$\textbf{(D)}$  $\Delta w$, measured in the region indicated by the dashed box in (C), when the axial load is increased from $F_a = 0.94 F_c$ to $F_a = 0.98 F_c$.
Circumferential $\textbf{(E)}$ and axial $\textbf{(F)}$ line-cuts through the center of the buckling mode (B) shown in black, the deformations (D) shown in pink, and originally imposed geometric imperfections (Fig.~\ref{F_1}D) shown in teal.
Note that the center of the numerically predicted buckling mode and measured deformations coincide, whereas the localized geometric imperfections are centered at a slightly different position.
}
\label{F_2}
\end{figure}

To develop an intuition for why \textit{ridge-tracking} works only when performed at the imparted imperfection, we compare numerical solutions to the von Karman-Donnell equations for the specific geometry of our modified sample with experimental measurements of the shell's deformations from axial loading. 
The modified shell's radial geometric imperfections are measured using the surface scanner, as shown in Fig.~\ref{F_1}D, and incorporated into the numerical shell model.
Numerical modal analysis is performed by employing a finite difference method, as detailed in the supplementary text.
Computational techniques, widely utilized for studying resonant frequencies in structures, enable calculating the energy cost of infinitesimal deviations from the equilibrium deformations [Supplementary Information].
The energy cost of deformations in the shape of a resonant mode, or eigenmode, is given by its eigenvalue; a buckling mode is an eigenmode associated with an eigenvalue of zero.
When no load is applied ($F_a = 0$), the model's eigenmode associated with the smallest eigenvalue, i.e. the most energetically favorable infinitesimal deformation, is delocalized with a periodic shape reminiscent of the checkerboard buckling modes predicted for perfect cylinders, as shown in Fig.~\ref{F_2}A.
However, when $F_a$ is set to the numerically calculated critical load, $F_{c, \text{sim}}$ (which differs from $F_c$ by $\approx 12\%$), the analysis shows a single localized buckling mode centered at the imparted defect, as shown in Fig.~\ref{F_2}B.
Notably, at the onset of buckling, the most energetically favorable deformation is one whose shape is qualitatively similar to the localized deformations imposed by the poker \cite{kriegesmannEffectsGeometricLoading2012}.

The experimental characterization of the modified shell's pre-buckling deformations involves determining relative radial deformations, $\Delta w = w(F_{a_1}) - w(F_{a_2})$, by subtracting surface scans of the shell taken at two axial loads, $F_{a_1}$ and $F_{a_2}$  [Supplementary Information].
The relative radial pre-buckling deformations resulting from increasing $F_a$ from $0.52 F_c$ to $0.71 F_c$, loads far from the buckling load, reveals a localized pattern centered on the imparted imperfection, as shown in Fig.~\ref{F_2}C, with a shape qualitatively resembling the numerically calculated buckling mode.
At loads closer to $F_c$, this similarity becomes more pronounced: increasing $F_a$ from $0.94 F_c$ to $0.98 F_c$ results in relative radial deformations whose shape and size are nearly identical to that of the numerically predicted buckling mode, as shown in Fig.~\ref{F_2}D.
Specifically, line-cuts through the center of the calculated mode are distinct in shape from the underlying imperfections and are in agreement with the experimentally observed deformations without the use of any fitting parameters, as shown in Fig.~\ref{F_2}E and F.
Subsequent high-speed videography confirms that failure initiates at the imparted imperfection, on which both the pre-buckling deformations and the numerically calculated buckling mode are centered (Movie S1).

\subsection*{Buckling of Generic Imperfect Shells}\label{generic}

\begin{figure}[h!]
\centering\includegraphics[width=0.9\textwidth]{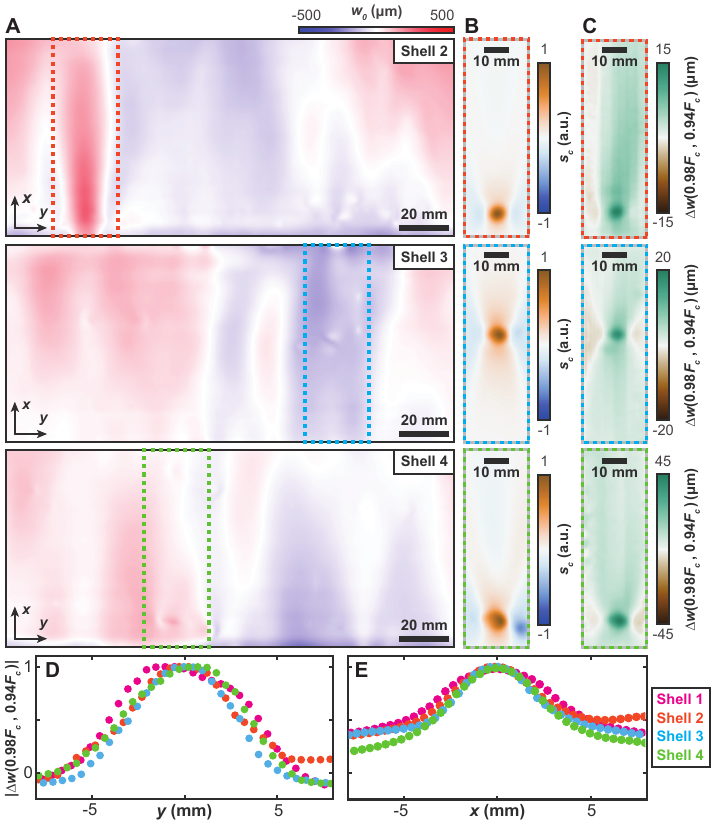}
\caption{\textbf{Localized buckling modes of unmodified cylindrical shells.}
$\textbf{(A)}$ Measured geometric imperfections, $w_0$, of three unmodified commercially manufactured shells. 
$\textbf{(B)}$ Numerically predicted localized buckling mode for shells with geometry shown in (A), presented for the region indicated by the dashed rectangular boxes.
$\textbf{(C)}$ Deformations, $\Delta w$, measured experimentally in the same region when the axial load is increased from $F_a = 0.94 F_c$ to $F_a = 0.98 F_c$.
Circumferential $\textbf{(D)}$ and axial $\textbf{(E)}$ line cuts of $\Delta w$, normalized by its maximum values, through the center of the deformations presented in (C) and Fig.~\ref{F_2}D.
}
\label{F_3}
\end{figure}

\begin{figure}[h!]
\centering\includegraphics[width=0.9\textwidth]{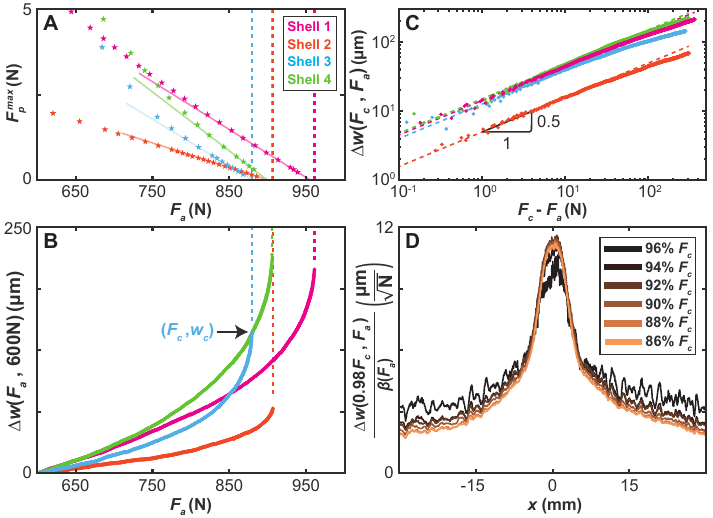}
\caption{\textbf{Real shells buckle via saddle-node bifurcation instability.}
$\textbf{(A)}$ $F^{\text{max}}_{p}$ as a function of $F_a$ generated by probing at the center of the predicted localized buckling mode.
Four successful ridge tracking procedures are shown resulting in accurate prediction of $F_c$ of both the modified and unmodified shells.
$\textbf{(B)}$ $\Delta w(F_a, 600 \text{N})$ experimentally measured at the center of the localized buckling mode as a function of $F_a$.
$\textbf{(C)}$ $\Delta w(F_c, F_a)$ as a function of $F_c - F_a$, extracted from the data in (B).
Dashed lines are fits to a function of the form $\Delta w(F_c, F_a) \propto (F_c - F_a)^{1/2}$.
$\textbf{(D)}$ Axial line-cuts of $\Delta w(0.98 F_c, F_a)$ through the center of the localized mode normalized by $\beta(F_a) = \sqrt{(F_c - F_a)} - \sqrt{(F_c - 0.98F_c)}$ for the shell presented in Fig.~\ref{F_1} and Fig.~\ref{F_2}.
}\label{F_4}
\end{figure}

In practice, commercially manufactured shells contain imperfections of varying types, shapes, and sizes, rarely exhibiting a single dominant localized defect.
This prompts the question: how does the buckling process differ for real, \textit{unmodified} shells?
To address this, we investigate the buckling behavior of three generic commercial shells whose inherent defects do not contain obviously dominant protuberances, as shown in Fig.~\ref{F_3}A.
For $F_a = 0$, simulations that incorporate each shell's unique initial shape once again indicate that the shells' eigenmode associated with the smallest eigenvalue is periodic (Ext. Fig.~\ref{S_8}). 
However, at $F_a = F_{c, \text{sim}}$, numerical analysis indicates that each shell has a unique localized buckling mode, as shown in Fig.~\ref{F_3}B.
The center of each calculated mode is used to predict where the post-buckling dynamics initiate and, thus, is used as the location at which to laterally probe the shell \cite{cucciaHitting2023}.
For all three unmodified shells, \textit{ridge-tracking} accurately predicts the experimentally observed buckling load to within 1\%, as shown in Fig.~\ref{F_4}A.

The reliability of the predictions enables us to safely approach the buckling load of an individual unmodified shell, providing the opportunity to experimentally resolve the deformations preceding the initiation of buckling in a realistic structure.
For each shell, the deformations resulting from applied axial loads approaching $F_c$, as shown in Fig.~\ref{F_3}C, are again strikingly similar to their respective numerically calculated localized buckling modes (Fig.~\ref{F_3}B).
When $F_a$ is increased past $F_c$, high-speed videography shows that failure initiates locally at the predicted location (Movie S2, Movie S3, and Movie S4).

The shape of each shell's localized buckling mode is similar across our samples despite variations of the underlying imperfections, as shown in Fig.~\ref{F_3}D and E.
Moreover, preliminary numerical analyses suggest that a single localized buckling mode is the defining feature of failure in realistically imperfect cylindrical shells. 
Notably, simulations show that the width of the localized buckling mode scales with the shell's radius, $r$, and thickness, $t$, as $\sqrt{\frac{r}{t}}$ (Ext.\ Fig.~\ref{S_5}), consistent with the scaling for the half-wavelength of the periodic buckling modes predicted by classical linear stability theory for mathematically perfect shells \cite{timoshenkoTheory1972}.

In cylindrical shells, the buckling mode is classically predicted to spontaneously emerge at the onset of buckling \cite{koiterEffect1963, timoshenkoTheory1972, brushBuckling1975, singerBuckling2001, hutchinsonPostbucklingTheory1970}; for our shells, the localized buckling mode experimentally manifests in the shell's pre-buckling deformations, well-below the buckling load.
We investigate the evolution of each shell's surface prior to buckling by experimentally measuring the shell's cumulative deformation resulting from quasi-statically increasing $F_a$ from $600\text{N}$ to $F_c$.
The amplitude of the radial deformations at the center of each shell's predicted mode increases smoothly and rapidly as $F_a \rightarrow F_c$, with a rate of change diverging when $F_a = F_c$, as shown in Fig.~\ref{F_4}B.
For all shells investigated, the deformation measurements display a $\Delta w(F_c, F_a) \propto (F_c - F_a)^{1/2}$ power-law scaling for loads spanning over two orders of magnitude, as shown in Fig.\ref{F_4}C.
The deformations along line-cuts through the failure initiation site, between $F_a = 0.98 F_c$ and variable lower loads, collapse to a universal shape resembling the calculated buckling mode when normalized by the load-dependent factor predicted from the $1/2$ power law, as shown in Fig.~\ref{F_4}D and Ext.\ Fig.~\ref{S_7}.

\subsection*{Buckling as a Saddle-Node Bifurcation Instability}\label{saddle}

In dynamical systems, the inherent nature of a bifurcation instability is reflected in the behavior of the system near a critical point; for the buckling of structures, a critical point arises when $F_a$ reaches the critical load, $F_c$.
As a system nears a critical point, its behavior is dominated by its eigenmodes with an eigenvalue of zero, which provide a reduced-dimensional basis for presenting and classifying universal features of the instability.
For a mechanical structure with a single, non-degenerate buckling mode, the onset of the instability is fully captured by the one-dimensional relationship between the applied load, $F_a$, and the deformation amplitudes in the shape of the buckling mode, $A_b$ \cite{lachmannMeasuring2023}.
Koiter's analytical solutions \cite{koiterEffect1963} to shell buckling point to a single stable equilibrium prior to buckling for which $A_b=0$.
When $F_a$ reaches $F_c$, the equilibrium path splits into multiple unstable branches, where $|A_b|$ spontaneously increases, marking the onset of buckling via a \emph{sub-critical pitchfork bifurcation}.

Our observation that the pre-buckling deformations of real shells closely mirror their localized buckling modes implies behavior known in shell buckling literature as a \emph{limit point} \cite{thompsonBasic1963, hutchinsonPostbucklingTheory1970} and in bifurcation theory as a \emph{saddle-node bifurcation} \cite{crawfordIntroduction1991}.
In a saddle-node bifurcation instability, the pre-buckling equilibrium amplitude, $A_b$, smoothly connects to the unstable post-buckling state with universal asymptotic behavior.
In the case of real shells, the equilibrium path, traditionally associated with a highly imperfection-sensitive non-linear energy minimization problem, has a simple one-dimensional relationship near $F_c$ given by
\begin{align}
A_b(F_c) - A_b(F_a) \propto w(F_c) - w(F_a) \propto (F_c - F_a)^{1/2},
\end{align}
in agreement with the experimentally observed $1/2$ power law scaling for the deformations at the buckling initiation site (Fig. \ref{F_4}C).
Thus, our results confirm that the pre-buckling deformations and subsequent post-buckling dynamics result from a \emph{saddle-node bifurcation} instability associated with a localized critical buckling mode.

This bifurcation framework for buckling of imperfect shells explains why \textit{ridge-tracking}, a procedure devised for perfect shells, applies to real structures.
A shell's buckling load is defined as the lowest load for which deformations in the shape of the buckling mode can grow without an associated energy cost.
\textit{Ridge-tracking}, which extrapolates when the probe force vanishes, predicts the load for which the deformations induced by poking can grow without energetic cost.
Thus, the load predicted by \textit{ridge-tracking} corresponds to the buckling load when the induced deformations are in the shape of the buckling mode.
We showed that real shells exhibit a unique localized buckling mode.
Therefore, because the localized deformations induced by poking \cite{kriegesmannEffectsGeometricLoading2012} have a shape similar to a localized buckling mode, \textit{ridge-tracking} will predict a real shell's buckling load if, and only if, probing is done at the center of this mode.
Furthermore, the empirically observed linear decrease in maximal lateral probing force with axial load (Fig. \ref{F_4}A) is theoretically predicted for a \emph{saddle-node bifurcation} when the induced deformations are in the shape of the buckling mode (supplementary text).

\subsection*{Conclusion}\label{Conc}

All naturally occurring and commercially produced shells are inherently imperfect and feature a broad distribution of defects.
Our findings reveal that buckling, despite being extremely sensitive to imperfections, is fully described by a simple and universal mechanism.
When a shell is subjected to loading, it develops small inhomogeneous surface deformations \textit{prior to} buckling that greatly influence its ultimate buckling dynamics.
Specifically, the buckling mode localizes at a single position determined by the background imperfections but with a shape that is largely insensitive to the underlying defect structure.
The buckling instability is characterized by a \emph{saddle-node bifurcation}, which prescribes the direct relationship between the pre-buckling deformations and the shell's buckling dynamics.
These results, shown for cylindrical shells, should be even more pronounced in spherical shells, where for perfect spheres a dimple-shaped imperfection has been shown to be as damaging at the ideal periodic  \cite{gerasimidisEstablishingBucklingKnockdowns2018, hutchinsonBucklingSphericalShells2016}.

This mechanistic understanding directly informs the development of methodologies for predicting generic unmodified shell structures' buckling properties, with far-reaching implications for safety and reliability.
Existing methods, notable examples being vibration-correlation \cite{kalninsExperimentalNondestructiveTest2015} and lateral probing techniques \cite{thompsonAdvances2015}, have physically well-motivated protocols applicable to cases where delocalized buckling modes govern failure, or when the location of buckling initiation is known a priori.
Evidence of pre-buckling localization provides realistic avenues for the improvement of existing methods, as successfully demonstrated here for lateral probing techniques, by accounting for the position of an individual shell's localized buckling mode.
However, using numerical methods to predict the mode's location, as done in this work, requires extremely accurate characterizations of defects and, even then, is very sensitive to boundary conditions \cite{schultzTest2018, hilburgerBuckling2008}.

Fortuitously, signatures of the buckling mode in the pre-buckling deformations, prescribed by the \emph{saddle-node bifurcation} instability, can instead be used to directly identify its center \cite{sunDigitalImageCorrelationaided2022}.
For commercial aluminum cans, localized deformations are clearly discernible at loads as low as $0.6 F_c$ (Ext.\ Fig.~\ref{S_6}) with amplitudes substantial enough for real-time detection and monitoring with commercially available sensors.
Furthermore, the universal power-law scaling that describes the critical behavior of pre-buckling deformations suggests that proximity to buckling may be estimated non-destructively using only radial deformation measurements.
Consequently, these results pave the way for the development of non-invasive stability monitoring systems that can sound the alarm when an increased load, be it from strong winds or excessive weight, would lead to catastrophic failure of crucial infrastructure.

\newpage
\null
\vfill
\vfill
\null

\newpage
\null
\vfill
\vfill
\null

\newpage
\null
\vfill
\vfill
\null

\newpage
\null
\vfill
\vfill
\null

\clearpage

\bibliography{Shells}

\clearpage

\include{Methods}

\section*{Acknowledgments}
This work was supported by the Israel Science Foundation (grant no. 2987/21). S.M.R. and N.L.C. acknowledge support from the Google Faculty Research Awards (2019). N.L.C. acknowledges support from the Harvard Porthcawl Innovation Fund and the Israeli Council of Higher Education. M.S. was partly supported by the Zuckerman Scholars Program while finalizing the manuscript. We thank John H.\ Hutchinson, Emmanuel Virot, and Tobias Schneider for helpful discussions, as well as Itamar Kolvin and Zvonimir Dogic for reading and commenting on the manuscript.

\subsection*{Author contributions}
M.S., N.L.C., and S.M.R. conceived the project, with M.S. providing the initial idea and conceptual framework. N.L.C. and S.M.R. designed the experimental system.  N.L.C. built the experimental system and obtained the experimental measurements. S.L. and M.S. assisted with data collection and high-speed photography. M.S. developed the finite difference numerical solver for the von Kármán-Donnell equations and conducted the numerical analysis. N.L.C., K.K.Y., and S.G. carried out the finite element simulations with Abaqus. M.S. and N.L.C. analyzed the results. M.S., N.L.C., and S.M.R. wrote the manuscript. S.G. and S.M.R. supervised the research. All authors discussed the results and commented on the manuscript.

\subsection*{Competing interests}
The authors declare no competing interests.

\subsection*{Data Availability}
The data supporting this study's findings are available in the Dryad repository at.

\clearpage

\import{./}{SI_Text}

\clearpage

\include{SuppData}

\end{document}

%% file: Methods.tex
\section*{Experimental Methods}

As model shells, we use empty aluminum cylindrical soda cans (7.5 fl oz) with length, $l=107$~mm, average radius, $r=28.6$~mm, and average thickness, $t\approx 80 \mu$m.
The complete experimental setup has been described in detail previously \cite{yadavNondestructive2021, cucciaHitting2023}.
A shell is placed upright in a custom tensile tester \cite{yadavNondestructive2021, cucciaHitting2023}, as illustrated schematically in Fig.~\ref{F_1}A. 
A vertical actuator with an attached load-cell compresses the shell between two parallel plates, exerting an axial load, $F_a$. 
At $F_a$ below the critical buckling load, $F_c$, the shell is stable and deforms elastically.
To characterize the shell's shape under different loading conditions, the setup is equipped with a chromatic confocal distance sensor (Marposs STIL OP-1000) capable of resolving 300~nm variations and 4.4~$\mu$m spot size (supplementary text).
The stability of the compressed shell is investigated by locally probing from the side with a lateral poker \cite{virotStability2017, abbasiProbing2021, yadavNondestructive2021, cucciaHitting2023} consisting of a $3.15$~mm aluminum ball-bearing attached to a load-cell moved by a horizontal actuator. 
When $F_a$ is increased past $F_c$, the shell buckles: an instability characterized by rapidly growing deformations initiates locally \cite{almrothExperimental1964, huhneNewApproachRobust2005, singerBuckling2002, cucciaHitting2023} and results in irreversible plastic damage.
To identify the location where failure initiates, footage of the shell's surface is taken at the onset of buckling with a high-speed camera (Phantom v711).

\subsection*{Bi-Axial Mechanical Tester}
A customized bi-axial mechanical tester (Instron, Inc.) is used to axially compress commercial aluminum cans, probe them from the side, and measure their surface deformations. 
Axial loads, $F_a$, are applied by compressing samples between two parallel loading plates.
Both loading plates are attached to rotary actuators (Oriental DGM130R-AZACR) that enable synchronous rotation about the axis of compression.
The distance between the top plate and the bottom plate is controlled by a vertical actuator and the top plate is equipped with a load cell (FUTEK LCB200) that directly reads out $F_a$.
The end shortening of the shell, and therefore $F_a$, is set by changing the distance between the loading plates. 
In practice, the separation is reduced at a constant speed of 5 mm per minute until the desired load is measured from the load cell, at which point the end shortening is held fixed.

Lateral probing measurements are performed using a blunt poker with a tip consisting of a 3.15~mm diameter aluminum ball bearing. 
The poker is attached to a lateral actuator (Thorlabs MTS25-Z8), pointing radially inwards on the shell.
The actuator sets the radial displacement, $D_p$, while the resulting lateral load, $F_p$, is measured from an attached load cell (FUTEK LSB200).
The poker and its lateral actuator are mounted onto an additional vertical actuator (Thorlabs LT300) that enables controlling the height of the probing location on the shell.
The circumferential probing location is changed by rotating the shell prior to axial loading. 
Both axial and lateral loading are always displacement controlled, and lateral poking is performed at set end shortening; the axial load varies by less than 2\% during poking.

The equilibrium surface deformations of a shell held at a fixed end displacement are measured using a chromatic confocal distance sensor (Marposs STIL OP-1000) whose position is controlled with both a vertical and horizontal linear actuator.
The chromatic confocal distance sensor has a working range of 2~mm and is capable of resolving 300~nm variations for a 500~$\mu$s integration time and 4.4~$\mu$m spot size.
Non-contact profilometry is performed by rastering the confocal distance sensor above the shell's surface using a combination of its linear actuators and the parallel loading plates' rotary actuators.
Efficient non-contact profilometry and autonomous lateral probing procedures are enabled by using an acquisition card (National Instrument DAQ USB-6001) to synchronously record measurements while controlling the position of various actuators.

Each sample's failure load, $F_c$, is found by continuously reducing the separation of the parallel plates until a precipitous drop in $F_a$ is measured.
Throughout the process, a high-speed camera (Phantom v711) is pointed at an array of mirrors arranged to show the shell's full height and circumference. 
Images are continuously acquired into a circular buffer, which is saved when the drop in $F_a$ is measured. 
These images are used to visualize the initiation of buckling. 

Our samples consist of 7.5 oz mini soda cans: aluminum cylindrical shells with an average radius $r = 28.6 mm$ and height $L = 107$~mm.
The soda cans were depressurized, emptied, and rinsed with water prior to testing.
During our experiments, we observed that the logos painted on the soda cans diminished the chromatic confocal distance sensor's effective working range.
Therefore, we employed a two-step procedure to remove the paint without roughening the shell's surface.
First, the shell was autoclaved on a dry cycle for 30 minutes to eliminate a protective polymer coating atop the paint.
Subsequently, the surface was gently washed with acetone to remove the commercial paint, resulting in a clean, mirror-like surface suitable for chromatic confocal scanning.
Throughout the entire sample preparation process, the shells were handled carefully to avoid imparting new defects.

\subsection*{Non-Contact Profilometry Rastering Methodologies}

Non-contact profilometry is performed by rastering a chromatic confocal distance sensor (CCDS) above the shell's surface.
We employ two different rastering methodologies, each with its own advantages and disadvantages, as described in this section.

In the first rastering methodology, the CCDS is moved vertically with a linear stage (Thorlabs LTS300) while the shell itself is rotated by synchronously rotating the parallel loading plates, as illustrated in Ext.\ Fig.~\ref{S_1}A.
At every vertical position, the shell is rotated a full 360 degrees while continuously recording distance measurements from the CCDS.
Repeating this process while changing the CCDS's vertical position yields a wide field-of-view distance profile as a function of the shell's height, $x$, and circumferential position, $y$ (see coordinate system defined in Ext.\ Fig.~\ref{S_1}A).
Since the CCDS points radially inwards (in the $\hat{z}$ direction) for the entire process, the distance measurements accurately characterize variations in the radial position of the shell's surface.
Measurements of unloaded shells, therefore, provide direct measurements of the radial geometric imperfections, $w_0(x,y)$, up to a constant offset.
Similarly, differences in distance profiles taken at different loads are measurements of the induced radial deformations, $\Delta w(F_1, F_2)(x,y)$.
As a result of minor misalignments, the rotation of the parallel loading plates causes variations in applied load, $F_a$, of up to $~5\%$.
Therefore, we do not use this rastering methodology when $F_a$ is close to $F_c$, as the load variations risk triggering a buckling event.

In the second rastering methodology, the CCDS moves both vertically and horizontally using linear stages, as illustrated in Ext.\ Fig.~\ref{S_1}B.
This confers the advantage that the shell remains perfectly stationary for the entire process and, therefore, the applied axial load is stable.
However, because the CCDS's position is controlled by linear stages, it moves along a plane tangent to the cylindrical shell's surface, providing a distance profile as a function of $x'$ and $y'$ (see coordinate system defined in Ext.\ Fig.~\ref{S_1}B).
This has two consequences.
First, it limits the scanning range to a subsection of the shell in the $y'$ direction, with width limited by the working distance of the CCDS.
Second, the distance measurements are now performed along the $\hat{z'}$ direction and must therefore be related to the variations in the radial direction, $\hat{z}$.
Difference measurements, $\Delta w'(x', y')$, are converted to radial deformation measurements, $\Delta w(x,y)$, using the following formula derived assuming $\Delta w \sim \Delta w' << r$: 
\begin{align}
    \Delta w(x', y') &\approx \sqrt{(\Delta w')^2 + r^2 + 2 \Delta w' 
    \sqrt{r^2 - y'^2}} - r \\
    y &= r \sin^{-1}(y'/r) \\
    x &= x'
\end{align}

As stated, both methodologies assume that the cylindrical shell is perfectly centered and does not shift with axial loading.
In practice, we observe that the top loading plate shifts on the order of 20~$\mu$m, with magnitude changing as a function of load.
These result in small tilts in our deformation measurements that, when possible, we remove in post-processing.

\subsection*{Geometric Imperfection Characterization}

Our numerical simulations of shells incorporate two different kinds of shell geometric imperfections: spatial variations in their radius ($w_0(x,y)$) and their thickness ($t(x,y)$). 

Wide field-of-view scans of the unloaded shells provide direct measurements of $w_0(x,y)$ with sub-micron accuracy.
However, our experimental apparatus cannot access the top or bottom $\sim$~1 cm of the shell.
We numerically pad our $w_0(x,y)$ measurements to match the length of the simulated shell with that of the real shell.
The radius of each shell shrinks by $\sim 1-2$~mm over the unmeasured $\sim$cm, so we use $w_0(0, y) = w_0(L, y) = 1.5$~mm as a boundary condition for the numerical padding.
At each circumferential point $y_i$, we populate the unknown values of $w_0(x, y_i)$ using a third-order polynomial with coefficients chosen such that $w_0(0, y) = w_0(L, y) = 1.5$~mm, $\frac{\partial w_0(0,y)}{\partial x} = \frac{\partial w_0(L, y)}{\partial x} = 0$, and continuity is preserved with the last measured values and their first derivatives. 
The padded $w_0(x,y)$ data sets are then digitally smoothed using a fourth-order low-pass Butterworth filter with a cutoff frequency of $2\pi/1.2$mm for the modified shell and $2\pi/1.5$mm for all three unmodified shells.
The resulting $w_0(x,y)$ data sets and their numerical first and second derivatives are visually smooth as required to use the von Karman-Donnell equations.

The thickness profile of our shells was determined by cutting samples into subsections along $y$, weighing them, and relating their mass to their thickness by assuming a uniform mass density.
We assume that the shell thickness profile resulting from the soda can manufacturing process primarily varies in the axial direction.
We used a laser cutter (N.T.N. Tech LM-UV5) to cut thin rings along the length of three typical unmodified cans; the rings varied in height from 1.1mm to 4mm, with the shorter strips near the ends of the can.
The strips were weighed, and their thickness was estimated using 
\begin{align}
2 \pi r h(x) t(x) \rho = m(x)
\end{align}
with $r=28.6$~mm the nominal radius of the can, $\rho = 2.7$~g/cm$^3$ the mass density of aluminum, and where $h(x)$, $t(x)$ and $m(x)$ are the ring's height, average thickness, and mass centered at the axial position $x$.
We note that the last $\sim$~1 cm of the cans, the regions corresponding to their caps, were too thick to be cut, and their thickness was not measured.

The shell thickness profile reveals that cans are thickest near their ends and that the majority of their bulk has a relatively constant thickness of around 80~$\mu$m. 
Furthermore, the determined thickness profile is consistent across the three shells tested.
Therefore, we assume the thickness profile of the three measured cans is representative of all our soda cans.
All our simulations use the same thickness profile interpolated from the data set of the three cans, as shown in Ext.\ Fig.~\ref{S_2}.

\section*{Numerical Methods}

The simulations presented in the main text are numerical solutions to the von Karman-Donnell equations for cylindrical shells.
In this section, we provide a summary of the equations, boundary conditions, and numerical techniques used to solve them. 
% These equations are often quoted to be of limited accuracy when the geometric imperfections are comparable to the shell's thickness \cite{}.
% The cylindrical shells investigated here have geometric imperfections with amplitudes up to $\sim 500 \mu$m, several times their thickness of $\sim 80\mu$m.
We compare the numerical predictions to those made by the finite element analysis software ABAQUS, which is known to be accurate for shells with larger imperfections.
% The two predictions are in excellent agreement, suggesting that the analytically derived von-Karman Donnell equations are accurate for a larger regime of parameters than anticipated by the literature.

\subsection*{von Karman-Donnell Equations}

In the formulation used by the von Karman-Donnell equations, the physical parameters are fully defined by the displacements of points along the shell's middle surface \cite{brushBuckling1975, timoshenkoTheory1972}. 
Variables are indexed by an $(x,y)$ coordinate, where $x$ refers to the axial position and $y$ is the circumferential position.
For a shell of length $L$ and radius $r$, the axial position $x$ varies from $0$ to $L$, and the circumferential position $y$ varies from $0$ to $2\pi r$, where $y = 0$ and $y = 2\pi r$ are the same point.
Specifically, the $(x,y)$ coordinate refers to a point on the middle surface of the unloaded structure without imperfections: even when the structure deforms, the $(x,y)$ coordinate refers to the point on the structure that was originally at the specified position.
Separate variables capture the deformations; in the axial, circumferential, and radial directions, the deformations are denoted $u(x,y)$, $v(x,y)$, and $w(x,y)$, respectively.
If the shell is imperfect and has variations in its curvature when unloaded, these are reflected by radial deformations $w_0(x,y)$.
The shell's thickness is given by $t(x,y)$ and can, in general, vary spatially. 
The shell is taken to have a constant Poisson ratio of $\nu$ and Young's modulus of $E$. 
In the rest of this section, we use the shorthand, $w(x,y) = w$, $\frac{\partial w(x,y)}{\partial x} = w_x$ and $\frac{\partial^2 w(x,y)}{\partial x^2} = w_{xx}$, etc... 

The strain in the axial direction, $\epsilon_1$, circumferential direction, $\epsilon_2$, and shear strain $\gamma$ at the middle surface are given by  \cite{brushBuckling1975}:
\begin{align}
    \epsilon_1 &= u_x + \frac{1}{2}w_x^2 + w_x w_{0, x} \\ 
    \epsilon_2 &= v_y + \frac{w}{r} + \frac{1}{2}w_y^2 + w_y w_{0,y} \\
    \gamma &= u_y + v_x  + w_x w_y + w_{0,x} w_y + w_x w_{0,y} 
\end{align}
where we have chosen the sign convention that outwards radial deformations are positive.

The stress in the axial direction, $\sigma_1$, circumferential direction, $\sigma_2$, and shear stress, $\tau$, at the middle surface are given by:
\begin{align}
    \sigma_1 &= \frac{E}{1-\nu^2}\left(\epsilon_1 + \nu \epsilon_2 \right) \\
    \sigma_2 &= \frac{E}{1-\nu^2}\left(\epsilon_2 + \nu \epsilon_1 \right) \\
    \tau &= \frac{E}{2(1+\nu)}\gamma
\end{align}

The von-Karman Donnell equations are the set of partial differential equations which must be satisfied in order for the deformations to represent an equilibrium solution. 
Typically, they are presented for a perfect shell with $w_0 = 0$ and a constant uniform thickness.
Relaxing these assumptions results in the differential equations:
\begin{align}
\begin{split}
 \frac{E}{12(1-\nu^2)}\left(\frac{\partial^2}{\partial x^2}\left(t^3 (w_{xx} + \nu w_{yy})\right) + \frac{\partial^2}{\partial y^2}\left(t^3 (w_{yy} + \nu w_{xx})\right) +
2(1-\nu)\frac{\partial^2}{\partial x \partial y}\left(t^3 w_{xy}\right)\right) \\ +
 t \sigma_1 (w_{xx} + w_{0, xx}) + t \sigma_2 (w_{yy} + w_{0, yy}) + 2 t \tau (w_{xy} + w_{0, xy}) = 0  
 \end{split}
\end{align}
\begin{align}
\frac{\partial}{\partial x}(t\sigma_1) + \frac{\partial}{\partial y}(t\tau) = 0 \\
\frac{\partial}{\partial y}(t\sigma_2) + \frac{\partial}{\partial x}(t\tau) = 0 
\end{align}

where all the variables in the above equations are functions of both $x$ and $y$. 
When there are no radial geometric imperfections, i.e. $w_0 = 0$, and the thickness is perfectly uniform such that $t(x,y) = t_{u}$, S11 simplifies to the cleaner, more familiar expression: 
\begin{align}
     \frac{Et_{u}^3}{12(1-\nu^2)}(w_{xxxx} + 2w_{xxyy} + w_{yyyy}) +
 t_{u} \sigma_1 w_{xx} + t_{u}\sigma_2 w_{yy} + 2 t_{u} \tau w_{xy} = 0  
\end{align}

Traditionally, equations S12 and S13 are combined into a single equation by introducing a stress function \cite{brushBuckling1975, timoshenkoTheory1972}.
However, realistic boundary conditions are more challenging to implement in this formulation. 
For simplicity, we elect to solve the three equations directly, which are functions of the axial, circumferential, and radial displacement functions.

Our boundary conditions are chosen such that the end of the shell maintains complete contact with the parallel loading plates at points held fixed by friction.
The friction requirement implies that the radial and circumferential displacements are zero at the end of the cylinder.
The contact requirement implies that the axial displacement at the ends is constant in the circumferential direction and that the end of the shell remains normal to the loading plate.
The load on the shell is introduced by the value chosen for the end displacement, $\delta$, fixed by the parallel loading plates. 
The complete boundary conditions for a shell of length $L$ are written as: 
\begin{align}
    w(0, y) = w(L, y) &= 0 \\
    w_x(0, y) = w_x(L, y) &= 0 \\
    v(0, y) = v(L, y) &= 0 \\
    u(0, y) = 0 \text{ and }  u(&L, y) = \delta     
\end{align}

\subsection*{Numerical Solver}

The equations are solved on a discretized two-dimensional rectangular grid with uniform spacing in both the $x$ and $y$ direction, specified by $dx$ and $dy$. 
At each $(x,y)$ coordinate, the continuous derivatives are replaced by their finite difference approximation.
One-dimensional finite difference approximations of the derivatives are computed to arbitrary order, $N$, using the Bjork-Pereyra algorithm, where errors scale as $O(dx^{N})$ and $O(dy^{N})$.
Matrix representations of the two-dimensional derivative operators, e.g. $\frac{\partial^2}{\partial x \partial y}$, are computed from outer products of the matrix representations of the one-dimensional operators.
Whenever possible, the finite difference approximations are central difference operators.
Near the ends of the cylinder, where $x = 0, L$ and the points required for a central difference approximation do not exist in both directions, we compute new finite difference operators that use more points, such that the accuracy of the approximation remains of the same order.
The boundary conditions specified in the previous section are incorporated as additional constraints.

The discretization generates a non-linear system of equations whose solution approximates the shell's equilibrium deformations.
We numerically compute the solution using a modified Newton-Raphson method, where the Jacobian is calculated analytically. 
The Jacobian of the shell for the equilibrium deformations at a given end-displacement, or load, is used to calculate the shell's eigenmodes with the smallest eigenvalues. 
Specifically, they are calculated using standard numerical techniques from a representation of the Jacobian's inverse in the basis of its Krylov subspace.

The buckling load is determined by taking advantage of the fact that, when the structure is unstable, the Newton-Raphson method fails to converge.
We slowly and incrementally increase the load, introduced by increasing $\delta$, until a solution cannot be found. 
The largest load for which a stable solution can be found is a good approximation of the buckling load when the step size is small.
This is demonstrated by the fact that the eigenvalue of the lowest value eigenmode for the last equilibrium solution trends to zero as the step size is reduced, as shown for a representative example in Ext.\ Fig.~\ref{S_3}.
The buckling mode, reported in the main text, corresponds to the eigenmode with the lowest eigenvalue at the last load for which a stable equilibrium was computed.

For the simulations presented in the text, we use $dx \approx dy \approx 0.25 mm$, forming a rectangular grid with 500,000 points.
We use finite difference approximation of order $N=6$, as higher-order approximations yield almost identical results.
The code is written in Python and uses Intel's Parallel Direct Sparse Solver Interface (PARADISO) to solve sparse matrices whenever required.

\subsection*{Comparison with ABAQUS}

We compare our predictions made by numerically solving the von Karman Donnell equations to those made by ABAQUS, a sophisticated general-purpose finite element-based software package. 
Simulations in ABAQUS are performed using a mesh of $S4R$ elements with the same density and spacing as we used when solving the von Karman-Donnell equations. 
The boundary conditions at the top and bottom of the shell are defined by a rigid link connecting a central node to the nodes on the cylinder's rim, the procedure originally devised in Haynie et al.  \cite{haynieValidationLowerBoundEstimates2012}. 

The buckling loads predicted by ABAQUS, $F_{c, AB}$ are within $\sim 2$\% of those predicted from the von Karman-Donnell equations, $F_{c, KD}$ for all four shells presented in the main text, as shown in Table \ref{T_1}.
We note that both predictions deviate significantly, by up to 10\%, from the buckling loads measured experimentally. 

\begin{table}[]
    \centering
    \begin{tabular}{c|c|c|c|c}
           &  Shell 1  &  Shell 2  &  Shell 3 & Shell 4  \\
         \hline
         $F_{c, vKD}$ & 1085.95 N & 852.92 N & 1174.76 N & 1011.66 N \\
        $F_{c, AB}$ & 1064.82 N & 833.38 N & 1153.85 N & 994.41 N\\
        $\left(F_{c, vKD} -F_{c, AB}\right)/F_{c, vKD}$  & 1.94 \% & 2.29 \% & 1.78 \% & 1.71\% \\
    \end{tabular}
    \caption{Buckling loads predicted by both ABAQUS and the von Karman-Donnell equations for all four shells presented in the main text. The shells are numbered in the order that they are presented in the main text.}
    \label{T_1}
\end{table}

Deformations at any given load are qualitatively and quantitatively in excellent agreement, as shown in Ext.\ Fig.~\ref{S_4}.

%% file: SI_Text.tex
\section*{Supplementary for A Universal Localized Failure Mechanism for Real Cylindrical Shells}

\textbf{Nicholas L.\ Cuccia, Marec Serlin, Kshitij K.\ Yadav, Sagy Lachmann, Simos Gerasimidis, Shmuel M.\ Rubinstein} \\

\setcounter{section}{0}

\section{Extended Discussion}

\subsection{Critical Behavior of Cylindrical Shells}

In this section, we discuss the general stability theory presented by Thompson\cite{thompsonBasic1963} using the language of cylindrical shells and connect the results to those presented in the main text. 
Consider a generic shell, with or without geometric imperfections, at an applied load $F_a$ below the critical load $F_c$.
The state of the shell is described by the deformation vector $\vec{A}  = \{u(x,y), v(x,y), w(x,y)\}$ where, as before, $u(x,y)$, $v(x,y)$ and $w(x,y)$ are the deformations of the middle surface in the axial, circumferential, and radial directions respectively.
$\vec{A}$ represents an equilibrium configuration when it minimizes the energy of the shell, $E(F_a, \vec{A})$ or, equivalently, satisfies the von Karman-Donnell equations.
The eigenmodes of the von-Karman Donnell equations provide a complete basis for expressing the effects of perturbations, $\vec{\delta A}$, on the equilibrium deformations. 
We call the $i^{\text{th}}$ eigenmode $\vec{\Psi_i}$ and its associated eigenvalue $\lambda_i$, where the eigenmodes are normalized such that $\int dxdy \|\vec{\Psi_i}\|^2 = S$, with $S$ the surface area of the reference shell.
We can express $\vec{\delta A}$ in the basis of the eigenmodes as:
\begin{align}
    \vec{\delta A} = \sum_i \delta a_i \vec{\Psi_i} \hspace{7 mm} \text{where} \hspace{7 mm} \delta a_i = \int dxdy \left(\vec{\delta A}\cdot \vec{\Psi_i}\right)
\end{align}

The change in energy resulting from $\vec{\delta A}$, can be simply expressed to lowest order as:
\begin{align}
    E(F_a, \vec{A} + \vec{\delta A}) \approx E(F_a, \vec{A}) + S\left(\sum_i \frac{\lambda_i}{2} (\delta a_i)^2\right)
\end{align}

The physical meaning of the eigenvalue is clear from the above equation: it corresponds to the normalized energetic cost of creating deformations in the shape of its associated eigenmode.
All the eigenmodes of a shell in stable equilibrium must have eigenvalues that are non-zero and positive, $\lambda_i > 0$. This guarantees that $\vec{A}$ is a local energy minimum.

We are interested in the load dependence of deformations around loads near the buckling load.
The critical buckling load, $F_c$, is defined as the first load that has an eigenmode with an eigenvalue of zero, i.e., a critical buckling mode; let's call the index of this eigenmode $i = c$ and $\lambda_c = 0$.
Hereon, we concern ourselves only with the general scenario where imperfections do not obey any symmetries.
We assume that this results in a unique buckling mode and an energy function, $E(F_c - \delta F, \vec{A} + \vec{\delta A})$, where the lowest order terms of its series expansion all have non-zero coefficients.
The energy at loads below the buckling load given by $F_a = F_c - \delta F$, expressed in the eigenmode basis defined at $F_a = F_c$, to lowest order is given by:
\begin{align}
    E(F_c - \delta F, \vec{A} + \vec{\delta A}) = E(F_c, \vec{A}) + S \left(\sum_{i, j, k} \frac{\gamma_{i, j, k}}{6} \delta a_i \delta a_j \delta a_k + \sum_i \left( \frac{\lambda_i}{2} \delta a_i^2 - \alpha_i \delta a_i \delta F \right) \right)
\end{align}
where $\alpha_i$ and $\gamma_{i, j, k}$ are system specific non-zero constants and the coefficient $\gamma_{i, j, k}$ is the same for all permutations of $i, j, $ and $k$. 
We note that, in order for an equilibrium to exist for $F_a < F_c$, $\alpha_c$ and $\gamma_{c, c, c}$ are constrained to have the same sign. 

To lowest order in $\delta F$, the equilibrium deformations are given by:
\begin{align}
    \delta a_c &= \sqrt{\frac{2\alpha_c \delta F}{\gamma_{c, c, c}}} \\
    \delta a_{i \neq c} &= \delta F \frac{\alpha_i - \gamma_{i, c, c} \alpha_c/\gamma_{c, c, c}}{\lambda_i}
\end{align}

We see that as $F_a \rightarrow F_c$,  $\delta a_c$ decreases as $\sqrt{\delta F}$ in contrast to $\delta a_{i\neq c}$ which decreases linearly.
This means that, as $\delta F \rightarrow 0$, the deformations will be dominated by $\delta a_c$.
Furthermore, the predicted load dependence of the amplitude of deformations in the shape of the buckling mode is exactly the 1/2 power law observed experimentally.
In his work, Thompson refers to failure in this manner as \textit{snapping} \cite{thompsonBasic1963}.
In the modern engineering literature, the same phenomenon is referred to as a structure's \textit{limit point}\cite{hutchinsonPostbucklingTheory1970, weingartenvBucklingThinwalledCircular2020}.
In bifurcation theory, this behavior is coined a \textit{saddle-node bifurcation}\cite{crawfordIntroduction1991}.

The energy of the shell can easily be related to the force required to introduce deformations, as done by a lateral probe. 
The force, $F_{probe}$ required to impose a deformation in the shape of the buckling mode on the shell is simply given by:
\begin{align}
    F_{probe} = -\frac{\partial E(F_c - \delta F, \vec{A} + \delta a_c \vec{\Psi_c})}{\partial (\delta a_c)}
\end{align}
The \textit{ridge-tracking} procedure discussed in the main text uses the maximum of the probe force as a function of the applied load. 
We see that the maximum force is given by:
\begin{align}
    F_{probe, max} = S \alpha_c \delta F + \mathcal{O}((\delta F)^2)
\end{align}
Thus, the linear asymptotic behavior observed from probing real shells is also expected for shells exhibiting a saddle-node bifurcation.

\subsection{Effect of Radius, Thickness, and Imperfection Amplitude on the Buckling Mode}

Here we present a preliminary numerical investigation into the effects of a shell's radius ($r$), thickness ($t$), and imperfection amplitude on its eventual buckling mode.
The values of $r$ and $t$ used to label simulations performed in this section represent the mean radius and thickness in the bulk of the shell.
The defects used in this investigation are taken from the experimentally measured geometric imperfections, referred to here as $w_{0, meas}$, presented in the main text (Fig.~\ref{F_1} and Fig.~\ref{F_3}).
We vary the imperfection amplitudes by linearly rescaling $w_{0, meas}$ by a unit-less factor $\alpha$, such that $w_0 = \alpha w_{0, meas}$.

The analysis is simplified by noting that the three-dimensional data set corresponding to variations with $r$, $t$, and $\alpha$ can be fully captured by varying only $t/r$ and $\alpha/r$.
This is easily seen by re-expressing all variables with units of length as a ratio to the radius $r$, such that $\Tilde{x} = x/r$, $\Tilde{w_0} = w_0/r$, $\Tilde{t} = t/r$, etc...
When propagated through the von Karman-Donnell equations, one observes that the only remaining independent parameters are $\Tilde{t}$ and $\Tilde{\alpha}$; the explicit dependence on the radius $r$ vanishes.
Simply stated, the von Karman-Donnell equations make identical predictions, up to a uniform scaling factor, so long as the aspect ratio of the shell and its imperfections remains unchanged.

We investigate the effects of varying $r/t = 1/\Tilde{t}$ from 100 to 1000, a range of values commonly found in commercially manufactured and industrially employed shells.
For each value of $\Tilde{t}$, multiple simulations are performed iterating over a range of values for $\alpha$, starting with $\alpha_1 = 1$ and following with 
 $\alpha_{n+1} = \alpha_n /2$ for subsequent iterations. 
We continue to run new simulations with smaller values for $\alpha$ until the shell's knockdown factor, $\kappa$, rises above 0.85.
For each value of $\Tilde{t}$, this procedure yields a set of simulations corresponding to different imperfection amplitudes and, therefore, knockdown factors.
Simulations varying $\Tilde{t}$ and $\alpha$ in this manner are performed for all four measured imperfections patterns presented in the main text.

The numerically predicted buckling modes, for all combinations of $\Tilde{t}$, $\alpha$, and $w_{0,meas}$, are localized, even for knockdown factors approaching 1.
The overall shape of the localized buckling mode is very similar in all cases, with a length-scale varying significantly with $\Tilde{t}$, as shown in Fig.~\ref{S_5}A-C.
Given the consistency of the localized buckling mode's shape, any reasonable characterization of its center region's width can be meaningfully compared across simulations with varying $\Tilde{t}$ and $\alpha$.
We elect to use the full-width half maximum of the buckling mode in the circumferential direction, $\xi_y$, as a measurement of the localized buckling mode's length scale.
We find that $\xi_y$ depends strongly on $\Tilde{t}$ and weakly on the knockdown factor $\kappa$.
For a fixed knockdown factor of $\kappa \approx 0.8$, the full-width half maximum is very similar across all four imperfection patterns and varies linearly with $\sqrt{\Tilde{t}}$, as shown in Fig.~\ref{S_5}D. 
The wavelength of the periodic oscillations of a perfect shell's buckling mode exhibits the same dependence on $r$ and $t$.
Furthermore, we observe that $\xi_y$ increases slightly for decreasing values of $\kappa$ at a fixed value of $\Tilde{t}$, as shown in Fig.~\ref{S_5}E.
For all simulated shells, $\xi_y$ is between 1.5\% and 5\% of the shell's circumference, suggesting that the region of localization is sufficiently large that deformations would be measurable in real systems of all aspect ratios.

\clearpage

\section*{Supplementary Movie Captions}

\noindent \textbf{Movie S1: Buckling Initiation for Shell 1.} High-speed video showing that buckling initiates at the predicted location for Shell 1. 
\break

\noindent \textbf{Movie S2: Buckling Initiation for Shell 2.} High-speed video showing that buckling initiates at the predicted location for Shell 2.
\break

\noindent \textbf{Movie S3: Buckling Initiation for Shell 3.} High-speed video showing that buckling initiates at the predicted location for Shell 3.
\break

\noindent \textbf{Movie S4: Buckling Initiation for Shell 4.} High-speed video showing that buckling initiates at the predicted location for Shell 4.
\break

%% file: SuppData.tex
\section*{Extended Data}

\renewcommand{\thefigure}{\arabic{figure}}
\renewcommand{\figurename}{Ext.\ Fig.}
\setcounter{figure}{0}

\null
\vfill
\begin{figure}[h!]
\centering\includegraphics[width=4in]{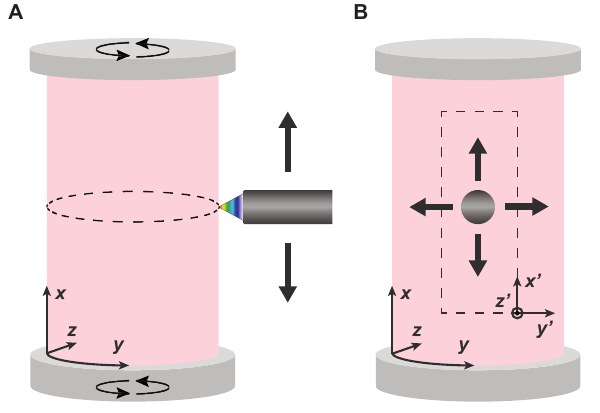}
\caption{\textbf{Schematic illustration of two different scanning methodologies.}
 $\textbf{(A)}$ The CCDS is moved vertically with a linear stage while the shell itself is rotated by synchronously rotating the parallel loading plates.
 This provides distance measurements as a function of $x$ and $y$ that are oriented in the $\hat{z}$ direction. 
$\textbf{(B)}$ The CCDS is moved vertically and horizontally with two linear stages while the shell itself is stationary. 
This provides distance measurements as a function of $x'$ and $y'$ that are oriented in the $\hat{z'}$ direction.
 }
\label{S_1}
\end{figure}
\vfill
\null

\newpage
\null
\vfill
\begin{figure}[h!]
\centering\includegraphics[width=2.3in]{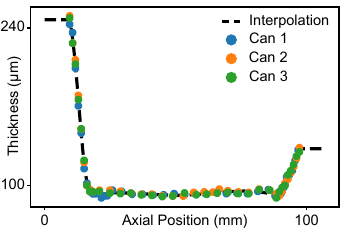}
\caption{Average thickness measurements as a function of axial position for three distinct soda cans. 
The black dashed line is the spline interpolation used as the thickness profile in all our numerical simulations.}
\label{S_2}
\end{figure}
\null
\vfill

\newpage
\null
\vfill
\begin{figure}[h!]
\centering\includegraphics[width=2.25in]{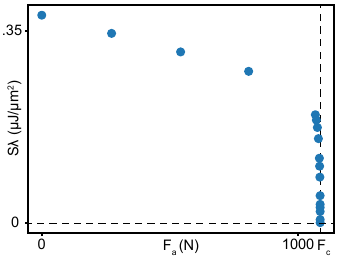}
\caption{To lowest order, the energy cost, $\Delta E$, of creating a deformation $w(x,y) = \delta a \Psi(x,y)$ is given by $\Delta E = \frac{1}{2}S\lambda \left(\delta a \right)^2$, where $\Psi(x,y)$ defines the shape of an eigenmode, $\lambda$ is its eigenvalue and $S$ is the shell's surface area (see extended discussion).
The coefficient $S\lambda$ captures the energy cost of creating deformations with intuitive units. 
The numerically predicted value for $S\lambda$ for the eigenmode with the smallest eigenvalue as a function of applied load, $F_a$, is presented for a shell with the geometric imperfections shown in Fig.~\ref{F_1}.
The energy cost rapidly trends to zero as $F_a$ nears the critical buckling load $F_c$.}
\label{S_3}
\end{figure}
\null
\vfill

\newpage
\null
\vfill
\begin{figure}[h!]
\centering\includegraphics[width=4.75in]{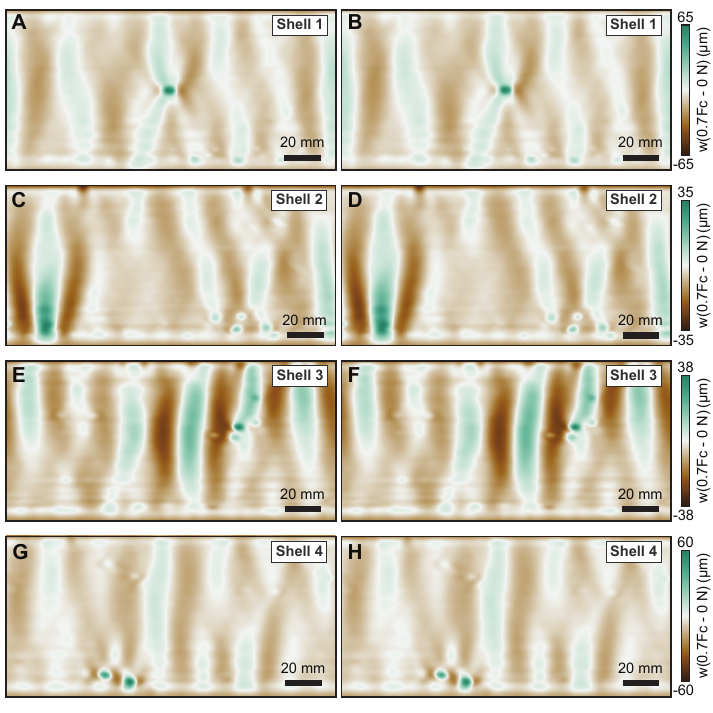}
\caption{\textbf{Comparison of the pre-buckling deformations predicted by ABAQUS and the von Karman-Donnell equations.}
Deformations between $F_a = 0.7F_c$ and $F_a = 0$ predicted numerically by the von Karman-Donnell equations (A, C, E, and G) and ABAQUS (B, D, F, H) for the shells with imperfection characterizations presented in the main text.}
\label{S_4}
\end{figure}
\null
\vfill

\newpage
\null
\vfill
\begin{figure}[h!]
\centering\includegraphics[width=5in]{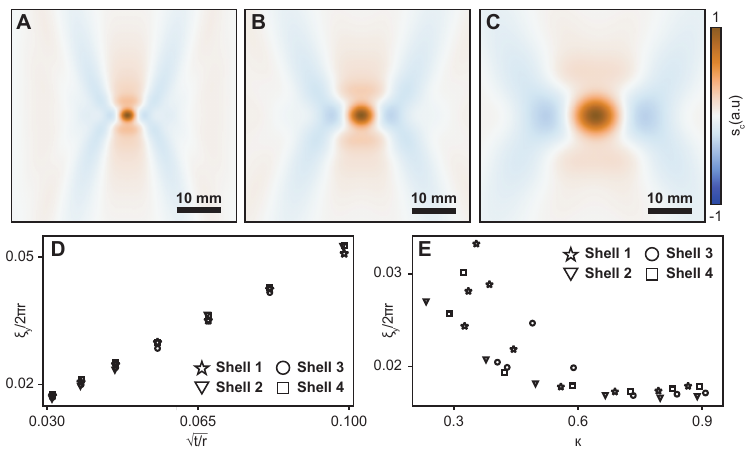}
\caption{\textbf{Effect of radius, thickness, and imperfections on the localized buckling mode.}
Buckling mode of a cylindrical shell with the geometric imperfection pattern of Shell 1 in the main text, with knockdown factor $\kappa \approx 0.8$, radius $r \approx 28.6 mm$ and $r/t$ set to $100$ (A), $500$ (B), and $1000$ (C). 
(D) Circumferential full-width half maximum of the localized buckling mode, $\xi_y$, as a function of $\Tilde{t}$ for $\kappa \approx 0.8$, where $\Tilde{t} = t/r$.  
The full-width half maximum of the numerically calculated buckling mode scales with $\sqrt{\frac{r}{t}}$. 
(E) $\xi_y$ as a function of the knockdown factor $\kappa$ for fixed ratio $r/t = 1000$. 
In both (D) and (E), the results from all four geometric imperfection patterns presented in the main text are displayed.
}
\label{S_5}
\end{figure}
\null
\vfill

\newpage
\null
\vfill
\begin{figure}[h!]
\centering\includegraphics[width=4.75in]{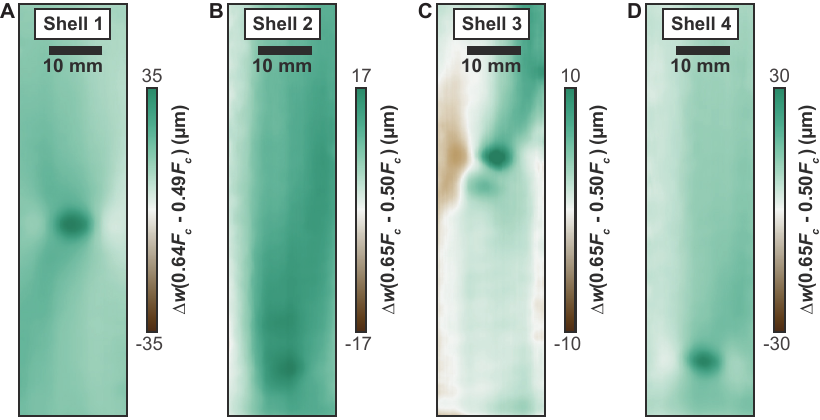}
\caption{\textbf{Localized Deformations at Axial Loads Well Below the Buckling Load.} Deformations, $\Delta w$, measured experimentally from $F_a =
0.49 F_c$ to $F_a = 0.64 F_c$ for Shell 1 (A). Deformations, $\Delta w$, measured experimentally from $F_a =
0.50 F_c$ to $F_a = 0.65 F_c$ for Shell 2 (B), Shell 3 (C), and Shell 4 (D).  These measured deformations coincide with the location of each shell's respective localized buckling mode and eventual failure initiation site.}
\label{S_6}
\end{figure}
\null
\vfill

\newpage
\null
\vfill
\begin{figure}[h!]
\centering\includegraphics[width=4.75in]{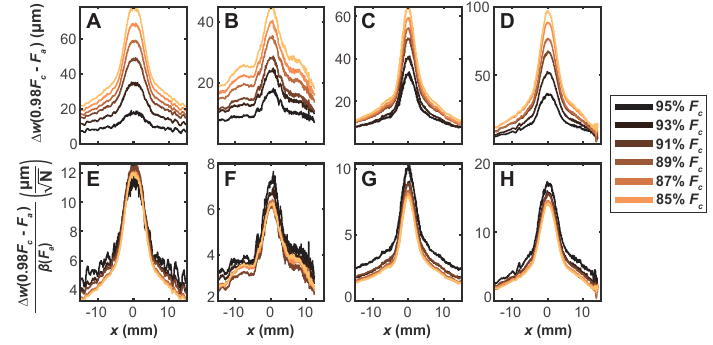}
\caption{\textbf{Axial Line Cuts for each Shell Collapse when Normalized} (A-D) Experimentally measured axial line-cuts of $\Delta w(0.98 F_c, F_a)$ through the center of the localized mode for each respective shell at various axial loads. (E-H) Line-cuts from (A-D) normalized by $\beta(F_a) = \sqrt{(F_c - F_a)} - \sqrt{(F_c - 0.98F_c)}$.  The line-cuts collapse for each shell across a wide range of axial loads. }
\label{S_7}
\end{figure}
\null
\vfill

\newpage
\null
\vfill
\begin{figure}[h!]
\centering\includegraphics[width=4.75in]{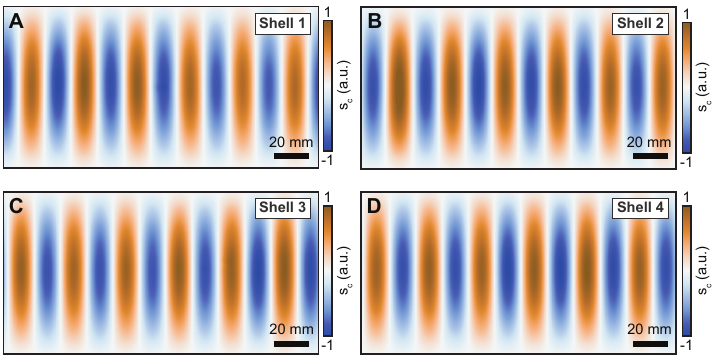}
\caption{\textbf{Eigenmode with Lowest Eigenvalue at $F_a = 0$} (A-D) Eigenmode with Lowest Eigenvalue at $F_a = 0$ for all four shells presented in the main text.}
\label{S_8}
\end{figure}

%% file: Buckling_Main.bbl
\begin{thebibliography}{10}
\expandafter\ifx\csname url\endcsname\relax
  \def\url#1{\burl{#1}}\fi
\expandafter\ifx\csname urlprefix\endcsname\relax\def\urlprefix{URL }\fi
\providecommand{\bibinfo}[2]{#2}
\providecommand{\eprint}[2][]{\url{#2}}
\providecommand{\doi}[1]{\url{https://doi.org/#1}}
\bibcommenthead

\bibitem{tsapisOnsetBucklingDrying2005}
\bibinfo{author}{Tsapis, N.} \emph{et~al.}
\newblock \bibinfo{title}{Onset of {{Buckling}} in {{Drying Droplets}} of {{Colloidal Suspensions}}}.
\newblock \emph{\bibinfo{journal}{Phys. Rev. Lett.}} \textbf{\bibinfo{volume}{94}}, \bibinfo{pages}{018302} (\bibinfo{year}{2005}).

\bibitem{sacannaLockKeyColloids2010}
\bibinfo{author}{Sacanna, S.}, \bibinfo{author}{Irvine, W. T.~M.}, \bibinfo{author}{Chaikin, P.~M.} \& \bibinfo{author}{Pine, D.~J.}
\newblock \bibinfo{title}{Lock and key colloids}.
\newblock \emph{\bibinfo{journal}{Nature}} \textbf{\bibinfo{volume}{464}}, \bibinfo{pages}{575--578} (\bibinfo{year}{2010}).

\bibitem{reisPerspectiveRevivalStructural2015}
\bibinfo{author}{Reis, P.~M.}
\newblock \bibinfo{title}{A {{Perspective}} on the {{Revival}} of {{Structural}} ({{In}}){{Stability With Novel Opportunities}} for {{Function}}: {{From Buckliphobia}} to {{Buckliphilia}}}.
\newblock \emph{\bibinfo{journal}{Journal of Applied Mechanics}} \textbf{\bibinfo{volume}{82}}, \bibinfo{pages}{111001} (\bibinfo{year}{2015}).

\bibitem{djellouliShell2024}
\bibinfo{author}{Djellouli, A.} \emph{et~al.}
\newblock \bibinfo{title}{Shell buckling for programmable metafluids}.
\newblock \emph{\bibinfo{journal}{Nature}} \textbf{\bibinfo{volume}{628}}, \bibinfo{pages}{545--550} (\bibinfo{year}{2024}).

\bibitem{weingartenvBucklingThinwalledCircular2020}
\bibinfo{author}{Weingarten, I., V}, \bibinfo{author}{Seide, P.} \& \bibinfo{author}{Peterson, P., J}.
\newblock \bibinfo{title}{Buckling of thin-walled circular cylinders}.
\newblock \bibinfo{type}{Tech. Rep.} \bibinfo{number}{NASA/SP-8007-2020/REV 2}, \bibinfo{institution}{{NASA}} (\bibinfo{year}{2020}).

\bibitem{thompsonAdvances2015}
\bibinfo{author}{Thompson, J. M.~T.}
\newblock \bibinfo{title}{Advances in {{Shell Buckling}}: {{Theory}} and {{Experiments}}}.
\newblock \emph{\bibinfo{journal}{Int. J. Bifurcation Chaos}} \textbf{\bibinfo{volume}{25}}, \bibinfo{pages}{1530001} (\bibinfo{year}{2015}).

\bibitem{elishakoffProbabilisticResolutionTwentieth2012}
\bibinfo{author}{Elishakoff, I.}
\newblock \bibinfo{title}{Probabilistic resolution of the twentieth century conundrum in elastic stability}.
\newblock \emph{\bibinfo{journal}{Thin-Walled Structures}} \textbf{\bibinfo{volume}{59}}, \bibinfo{pages}{35--57} (\bibinfo{year}{2012}).

\bibitem{ankalhopeNondestructivePredictionBuckling2022}
\bibinfo{author}{Ankalhope, S.} \& \bibinfo{author}{Jose, S.}
\newblock \bibinfo{title}{Non-destructive prediction of buckling load of axially compressed cylindrical shells using {{Least Resistance Path}} to {{Probing}}}.
\newblock \emph{\bibinfo{journal}{Thin-Walled Structures}} \textbf{\bibinfo{volume}{170}}, \bibinfo{pages}{108497} (\bibinfo{year}{2022}).

\bibitem{thompsonBasic1963}
\bibinfo{author}{Thompson, J.}
\newblock \bibinfo{title}{Basic principles in the general theory of elastic stability}.
\newblock \emph{\bibinfo{journal}{Journal of the Mechanics and Physics of Solids}} \textbf{\bibinfo{volume}{11}}, \bibinfo{pages}{13--20} (\bibinfo{year}{1963}).

\bibitem{timoshenkoTheory1972}
\bibinfo{author}{Timoshenko, S.} \& \bibinfo{author}{Gere, J.~M.}
\newblock \emph{\bibinfo{title}{Theory of Elastic Stability}} \bibinfo{edition}{2nd ed.} edn (\bibinfo{publisher}{{Dover Publications}}, \bibinfo{address}{{Mineola, N.Y}}, \bibinfo{year}{1972}).

\bibitem{brushBuckling1975}
\bibinfo{author}{Brush, D.~O.} \& \bibinfo{author}{Almroth, B.~O.}
\newblock \emph{\bibinfo{title}{Buckling of Bars, Plates, and Shells}}  (\bibinfo{publisher}{{McGraw-Hill}}, \bibinfo{address}{{New York}}, \bibinfo{year}{1975}).

\bibitem{singerBuckling2001}
\bibinfo{author}{Singer, J.}, \bibinfo{author}{Arbocz, J.} \& \bibinfo{author}{Weller, T.}
\newblock \emph{\bibinfo{title}{Buckling {{Experiments}}: {{Experimental Methods}} in {{Buckling}} of {{Thin-Walled Structures}}}} Vol.~\bibinfo{volume}{1} (\bibinfo{publisher}{{John Wiley \& Sons, Inc.}}, \bibinfo{address}{{New York}}, \bibinfo{year}{2001}).

\bibitem{koiterEffect1963}
\bibinfo{author}{Koiter, W.~T.}
\newblock \bibinfo{title}{The {{Effect}} of {{Axisymmetric Imperfections}} on the {{Buckling}} of {{Cylindrical Shells Under Axial Compression}}}.
\newblock \emph{\bibinfo{journal}{Koninklijke Nederlandse Akademie van Wetenschappen}} \textbf{\bibinfo{volume}{66}}, \bibinfo{pages}{265--279} (\bibinfo{year}{1963}).

\bibitem{hutchinsonPostbucklingTheory1970}
\bibinfo{author}{Hutchinson, J.~W.} \& \bibinfo{author}{Koiter, W.~T.}
\newblock \bibinfo{title}{Postbuckling {{Theory}}}.
\newblock \emph{\bibinfo{journal}{Applied Mechanics Reviews}} \textbf{\bibinfo{volume}{23}}, \bibinfo{pages}{1352--1366} (\bibinfo{year}{1970}).

\bibitem{almrothExperimental1964}
\bibinfo{author}{Almroth, B.~O.}, \bibinfo{author}{Holmes, A. M.~C.} \& \bibinfo{author}{Brush, D.~O.}
\newblock \bibinfo{title}{An experimental study of the bucking of cylinders under axial compression: {{Test}} program is aimed at determining the causes of discrepancy between theoretical and experimental results for the buckling load of axially compressed cylindrical shells}.
\newblock \emph{\bibinfo{journal}{Experimental Mechanics}} \textbf{\bibinfo{volume}{4}}, \bibinfo{pages}{263--270} (\bibinfo{year}{1964}).

\bibitem{tsienTheoryBucklingThin1942}
\bibinfo{author}{Tsien, H.-S.}
\newblock \bibinfo{title}{A {{Theory}} for the {{Buckling}} of {{Thin Shells}}}.
\newblock \emph{\bibinfo{journal}{Journal of the Aeronautical Sciences}} \textbf{\bibinfo{volume}{9}}, \bibinfo{pages}{373--384} (\bibinfo{year}{1942}).

\bibitem{leeGeometric2016}
\bibinfo{author}{Lee, A.}, \bibinfo{author}{L{\'o}pez~Jim{\'e}nez, F.}, \bibinfo{author}{Marthelot, J.}, \bibinfo{author}{Hutchinson, J.~W.} \& \bibinfo{author}{Reis, P.~M.}
\newblock \bibinfo{title}{The {{Geometric Role}} of {{Precisely Engineered Imperfections}} on the {{Critical Buckling Load}} of {{Spherical Elastic Shells}}}.
\newblock \emph{\bibinfo{journal}{Journal of Applied Mechanics}} \textbf{\bibinfo{volume}{83}}, \bibinfo{pages}{111005} (\bibinfo{year}{2016}).

\bibitem{singerBuckling2002}
\bibinfo{author}{Singer, J.}, \bibinfo{author}{Arbocz, J.} \& \bibinfo{author}{Weller, T.}
\newblock \emph{\bibinfo{title}{Buckling {{Experiments}}: {{Experimental Methods}} in {{Buckling}} of {{Thin-Walled Structures}}}} Vol.~\bibinfo{volume}{2} (\bibinfo{publisher}{{John Wiley \& Sons, Inc.}}, \bibinfo{address}{{Hoboken, NJ, USA}}, \bibinfo{year}{2002}).

\bibitem{huhneNewApproachRobust2005}
\bibinfo{author}{H{\"u}hne, C.}, \bibinfo{author}{Rolfes, R.} \& \bibinfo{author}{Tessmer, J.}
\newblock \emph{\bibinfo{title}{A new approach for robust design of composite cylindrical shells under axial compression}} (\bibinfo{address}{{Noordwijk, The Netherlands}}, \bibinfo{year}{2005}).

\bibitem{huhneRobust2008}
\bibinfo{author}{H{\"u}hne, C.}, \bibinfo{author}{Rolfes, R.}, \bibinfo{author}{Breitbach, E.} \& \bibinfo{author}{Te{\ss}mer, J.}
\newblock \bibinfo{title}{Robust design of composite cylindrical shells under axial compression \textemdash{} {{Simulation}} and validation}.
\newblock \emph{\bibinfo{journal}{Thin-Walled Structures}} \textbf{\bibinfo{volume}{46}}, \bibinfo{pages}{947--962} (\bibinfo{year}{2008}).

\bibitem{cucciaHitting2023}
\bibinfo{author}{Cuccia, N.~L.} \emph{et~al.}
\newblock \bibinfo{title}{Hitting the mark: Probing at the initiation site allows for accurate prediction of a thin shell's buckling load}.
\newblock \emph{\bibinfo{journal}{Phil. Trans. R. Soc. A.}} \textbf{\bibinfo{volume}{381}}, \bibinfo{pages}{20220036} (\bibinfo{year}{2023}).

\bibitem{hilburgerShell2006}
\bibinfo{author}{Hilburger, M.~W.}, \bibinfo{author}{Nemeth, M.~P.} \& \bibinfo{author}{Starnes, J.~H.}
\newblock \bibinfo{title}{Shell {{Buckling Design Criteria Based}} on {{Manufacturing Imperfection Signatures}}}.
\newblock \emph{\bibinfo{journal}{AIAA Journal}} \textbf{\bibinfo{volume}{44}}, \bibinfo{pages}{654--663} (\bibinfo{year}{2006}).

\bibitem{hilburgerBuckling2008}
\bibinfo{author}{Hilburger, M.~W.}
\newblock \bibinfo{title}{Buckling {{Test Results}} from the 8-{{Foot-Diameter Orthogrid-Stiffened Cylinder Test Article TA0}}} \bibinfo{pages}{72} (\bibinfo{year}{2008}).

\bibitem{kriegesmannEffectsGeometricLoading2012}
\bibinfo{author}{Kriegesmann, B.}, \bibinfo{author}{Hilburger, M.} \& \bibinfo{author}{Rolfes, R.}
\newblock \emph{\bibinfo{title}{The {{Effects}} of {{Geometric}} and {{Loading Imperfections}} on the {{Response}} and {{Lower-Bound Buckling Load}} of a {{Compression-Loaded Cylindrical Shell}}}}, \bibinfo{pages}{1--10} (\bibinfo{publisher}{{American Institute of Aeronautics and Astronautics}}, \bibinfo{address}{{Reston, Virigina}}, \bibinfo{year}{2012}).

\bibitem{kalninsExperimentalNondestructiveTest2015}
\bibinfo{author}{Kalnins, K.} \emph{et~al.}
\newblock \bibinfo{title}{Experimental {{Nondestructive Test}} for {{Estimation}} of {{Buckling Load}} on {{Unstiffened Cylindrical Shells Using Vibration Correlation Technique}}}.
\newblock \emph{\bibinfo{journal}{Shock and Vibration}} \textbf{\bibinfo{volume}{2015}}, \bibinfo{pages}{1--8} (\bibinfo{year}{2015}).

\bibitem{gerasimidisEstablishingBucklingKnockdowns2018}
\bibinfo{author}{Gerasimidis, S.}, \bibinfo{author}{Virot, E.}, \bibinfo{author}{Hutchinson, J.~W.} \& \bibinfo{author}{Rubinstein, S.~M.}
\newblock \bibinfo{title}{On establishing buckling knockdowns for imperfection-sensitive shell structures}.
\newblock \emph{\bibinfo{journal}{Journal of Applied Mechanics, Transactions ASME}} \textbf{\bibinfo{volume}{85}} (\bibinfo{year}{2018}).

\bibitem{donnellEffect1950}
\bibinfo{author}{Donnell, L.~H.} \& \bibinfo{author}{Wan, C.~C.}
\newblock \bibinfo{title}{Effect of {{Imperfections}} on {{Buckling}} of {{Thin Cylinders}} and {{Columns Under Axial Compression}}}.
\newblock \emph{\bibinfo{journal}{Journal of Applied Mechanics}} \textbf{\bibinfo{volume}{17}}, \bibinfo{pages}{73--83} (\bibinfo{year}{1950}).

\bibitem{horakCylinder2006}
\bibinfo{author}{Hor{\'a}k, J.}, \bibinfo{author}{Lord, G.~J.} \& \bibinfo{author}{Peletier, M.~A.}
\newblock \bibinfo{title}{Cylinder {{Buckling}}: {{The Mountain Pass}} as an {{Organizing Center}}}.
\newblock \emph{\bibinfo{journal}{SIAM J. Appl. Math.}} \textbf{\bibinfo{volume}{66}}, \bibinfo{pages}{1793--1824} (\bibinfo{year}{2006}).

\bibitem{gerasimidisDentImperfectionsShell2021}
\bibinfo{author}{Gerasimidis, S.} \& \bibinfo{author}{Hutchinson, J.~W.}
\newblock \bibinfo{title}{Dent {{Imperfections}} in {{Shell Buckling}}: {{The Role}} of {{Geometry}}, {{Residual Stress}}, and {{Plasticity}}}.
\newblock \emph{\bibinfo{journal}{Journal of Applied Mechanics, Transactions ASME}} \textbf{\bibinfo{volume}{88}}, \bibinfo{pages}{1--9} (\bibinfo{year}{2021}).

\bibitem{royerProbing2022}
\bibinfo{author}{Royer, F.} \& \bibinfo{author}{Pellegrino, S.}
\newblock \bibinfo{title}{Probing the {{Stability}} of {{Ladder-Type Coilable Space Structures}}}.
\newblock \emph{\bibinfo{journal}{AIAA Journal}} \textbf{\bibinfo{volume}{60}}, \bibinfo{pages}{2000--2012} (\bibinfo{year}{2022}).

\bibitem{royerExperimentally2023}
\bibinfo{author}{Royer, F.} \& \bibinfo{author}{Pellegrino, S.}
\newblock \bibinfo{title}{Experimentally probing the stability of thin-shell structures under pure bending}.
\newblock \emph{\bibinfo{journal}{Philosophical Transactions of the Royal Society A: Mathematical, Physical and Engineering Sciences}} \textbf{\bibinfo{volume}{381}}, \bibinfo{pages}{20220024} (\bibinfo{year}{2023}).

\bibitem{haynieValidationLowerBoundEstimates2012}
\bibinfo{author}{Haynie, W.}, \bibinfo{author}{Hilburger, M.}, \bibinfo{author}{Bogge, M.}, \bibinfo{author}{Maspoli, M.} \& \bibinfo{author}{Kriegesmann, B.}
\newblock \emph{\bibinfo{title}{Validation of {{Lower-Bound Estimates}} for {{Compression-Loaded Cylindrical Shells}}}} (\bibinfo{publisher}{{American Institute of Aeronautics and Astronautics}}, \bibinfo{address}{{Honolulu, Hawaii}}, \bibinfo{year}{2012}).

\bibitem{castroGeometricImperfectionsLowerbound2014}
\bibinfo{author}{Castro, S.~G.} \emph{et~al.}
\newblock \bibinfo{title}{Geometric imperfections and lower-bound methods used to calculate knock-down factors for axially compressed composite cylindrical shells}.
\newblock \emph{\bibinfo{journal}{Thin-Walled Structures}} \textbf{\bibinfo{volume}{74}}, \bibinfo{pages}{118--132} (\bibinfo{year}{2014}).

\bibitem{schultzTest2018}
\bibinfo{author}{Schultz, M.~R.} \emph{et~al.}
\newblock \emph{\bibinfo{title}{Test and {{Analysis}} of a {{Buckling-Critical Large-Scale Sandwich Composite Cylinder}}}} (\bibinfo{publisher}{{American Institute of Aeronautics and Astronautics}}, \bibinfo{address}{{Kissimmee, Florida}}, \bibinfo{year}{2018}).

\bibitem{kreilosFully2017}
\bibinfo{author}{Kreilos, T.} \& \bibinfo{author}{Schneider, T.~M.}
\newblock \bibinfo{title}{Fully localized post-buckling states of cylindrical shells under axial compression}.
\newblock \emph{\bibinfo{journal}{Proceedings of the Royal Society A: Mathematical, Physical and Engineering Sciences}} \textbf{\bibinfo{volume}{473}}, \bibinfo{pages}{20170177} (\bibinfo{year}{2017}).

\bibitem{grohRole2019}
\bibinfo{author}{Groh, {\relax RMJ}.} \& \bibinfo{author}{Pirrera, A.}
\newblock \bibinfo{title}{On the role of localizations in buckling of axially compressed cylinders}.
\newblock \emph{\bibinfo{journal}{Proceedings of the Royal Society A}} \textbf{\bibinfo{volume}{475}}, \bibinfo{pages}{20190006} (\bibinfo{year}{2019}).

\bibitem{audolyLocalization2020}
\bibinfo{author}{Audoly, B.} \& \bibinfo{author}{Hutchinson, J.~W.}
\newblock \bibinfo{title}{Localization in spherical shell buckling}.
\newblock \emph{\bibinfo{journal}{Journal of the Mechanics and Physics of Solids}} \textbf{\bibinfo{volume}{136}} (\bibinfo{year}{2020}).

\bibitem{sunDigitalImageCorrelationaided2022}
\bibinfo{author}{Sun, W.}, \bibinfo{author}{Zhu, T.}, \bibinfo{author}{Li, F.} \& \bibinfo{author}{Lin, G.}
\newblock \bibinfo{title}{Digital image correlation-aided non-destructive buckling load prediction of cylindrical shells}.
\newblock \emph{\bibinfo{journal}{International Journal of Solids and Structures}} \textbf{\bibinfo{volume}{254--255}}, \bibinfo{pages}{111941} (\bibinfo{year}{2022}).

\bibitem{yadavNondestructive2021}
\bibinfo{author}{Yadav, K.~K.}, \bibinfo{author}{Cuccia, N.~L.}, \bibinfo{author}{Virot, E.}, \bibinfo{author}{Rubinstein, S.~M.} \& \bibinfo{author}{Gerasimidis, S.}
\newblock \bibinfo{title}{A {{Nondestructive Technique}} for the {{Evaluation}} of {{Thin Cylindrical Shells}}' {{Axial Buckling Capacity}}}.
\newblock \emph{\bibinfo{journal}{Journal of Applied Mechanics, Transactions ASME}} \textbf{\bibinfo{volume}{88}} (\bibinfo{year}{2021}).

\bibitem{abramianNondestructive2020}
\bibinfo{author}{Abramian, A.}, \bibinfo{author}{Virot, E.}, \bibinfo{author}{Lozano, E.}, \bibinfo{author}{Rubinstein, S.~M.} \& \bibinfo{author}{Schneider, T.~M.}
\newblock \bibinfo{title}{Nondestructive {{Prediction}} of the {{Buckling Load}} of {{Imperfect Shells}}}.
\newblock \emph{\bibinfo{journal}{Physical Review Letters}} \textbf{\bibinfo{volume}{125}}, \bibinfo{pages}{225504} (\bibinfo{year}{2020}).

\bibitem{virotStability2017}
\bibinfo{author}{Virot, E.}, \bibinfo{author}{Kreilos, T.}, \bibinfo{author}{Schneider, T.~M.} \& \bibinfo{author}{Rubinstein, S.~M.}
\newblock \bibinfo{title}{Stability {{Landscape}} of {{Shell Buckling}}}.
\newblock \emph{\bibinfo{journal}{Physical Review Letters}} \textbf{\bibinfo{volume}{119}}, \bibinfo{pages}{1--5} (\bibinfo{year}{2017}).

\bibitem{yadavImperfectionInsensitiveThin2020}
\bibinfo{author}{Yadav, K.~K.} \& \bibinfo{author}{Gerasimidis, S.}
\newblock \bibinfo{title}{Imperfection insensitive thin cylindrical shells for next generation wind turbine towers}.
\newblock \emph{\bibinfo{journal}{Journal of Constructional Steel Research}} \textbf{\bibinfo{volume}{172}}, \bibinfo{pages}{106228} (\bibinfo{year}{2020}).

\bibitem{marthelotBuckling2017}
\bibinfo{author}{Marthelot, J.}, \bibinfo{author}{Jim{\'e}nez, F.~L.}, \bibinfo{author}{Lee, A.}, \bibinfo{author}{Hutchinson, J.~W.} \& \bibinfo{author}{Reis, P.~M.}
\newblock \bibinfo{title}{Buckling of a {{Pressurized Hemispherical Shell Subjected}} to a {{Probing Force}}}.
\newblock \emph{\bibinfo{journal}{Journal of Applied Mechanics, Transactions ASME}} \textbf{\bibinfo{volume}{84}}, \bibinfo{pages}{1--9} (\bibinfo{year}{2017}).

\bibitem{abbasiProbing2021}
\bibinfo{author}{Abbasi, A.}, \bibinfo{author}{Yan, D.} \& \bibinfo{author}{Reis, P.~M.}
\newblock \bibinfo{title}{Probing the buckling of pressurized spherical shells}.
\newblock \emph{\bibinfo{journal}{Journal of the Mechanics and Physics of Solids}} \textbf{\bibinfo{volume}{155}}, \bibinfo{pages}{104545} (\bibinfo{year}{2021}).

\bibitem{thompsonQuantified2014}
\bibinfo{author}{Thompson, J. M.~T.} \& \bibinfo{author}{{van der Heijden}, G. H.~M.}
\newblock \bibinfo{title}{Quantified "{{Shock-Sensitivity}}" {{Above}} the {{Maxwell Load}}}.
\newblock \emph{\bibinfo{journal}{Int. J. Bifurcation Chaos}} \textbf{\bibinfo{volume}{24}}, \bibinfo{pages}{1430009} (\bibinfo{year}{2014}).

\bibitem{thompsonShockSensitivity2016}
\bibinfo{author}{Thompson, J. M.~T.} \& \bibinfo{author}{Sieber, J.}
\newblock \bibinfo{title}{Shock-{{Sensitivity}} in {{Shell-Like Structures}}: {{With Simulations}} of {{Spherical Shell Buckling}}}.
\newblock \emph{\bibinfo{journal}{Int. J. Bifurcation Chaos}} \textbf{\bibinfo{volume}{26}}, \bibinfo{pages}{1630003} (\bibinfo{year}{2016}).

\bibitem{hutchinsonNonlinear2017}
\bibinfo{author}{Hutchinson, J.~W.}, \bibinfo{author}{Thompson, T.} \& \bibinfo{author}{Michael, J.}
\newblock \bibinfo{title}{Nonlinear buckling behaviour of spherical shells: {{Barriers}} \& symmetry-breaking dimples}.
\newblock \emph{\bibinfo{journal}{Philosophical Transactions of the Royal Society A: Mathematical, Physical and Engineering Sciences}} \textbf{\bibinfo{volume}{375}} (\bibinfo{year}{2017}).

\bibitem{hutchinsonImperfections2018}
\bibinfo{author}{Hutchinson, J.~W.} \& \bibinfo{author}{Thompson, J. M.~T.}
\newblock \bibinfo{title}{Imperfections and energy barriers in shell buckling}.
\newblock \emph{\bibinfo{journal}{International Journal of Solids and Structures}} \textbf{\bibinfo{volume}{148--149}}, \bibinfo{pages}{157--168} (\bibinfo{year}{2018}).

\bibitem{lachmannMeasuring2023}
\bibinfo{author}{Lachmann, S.} \& \bibinfo{author}{Rubinstein, S.~M.}
\newblock \bibinfo{title}{Measuring the energy landscape: An experimental approach to the study of buckling in thin shells}.
\newblock \emph{\bibinfo{journal}{Philosophical Transactions of the Royal Society A: Mathematical, Physical and Engineering Sciences}} \textbf{\bibinfo{volume}{381}}, \bibinfo{pages}{20220027} (\bibinfo{year}{2023}).

\bibitem{crawfordIntroduction1991}
\bibinfo{author}{Crawford, J.~D.}
\newblock \bibinfo{title}{Introduction to bifurcation theory}.
\newblock \emph{\bibinfo{journal}{Rev. Mod. Phys.}} \textbf{\bibinfo{volume}{63}}, \bibinfo{pages}{991--1037} (\bibinfo{year}{1991}).

\bibitem{hutchinsonBucklingSphericalShells2016}
\bibinfo{author}{Hutchinson, J.~W.}
\newblock \bibinfo{title}{Buckling of spherical shells revisited}.
\newblock \emph{\bibinfo{journal}{Proceedings of the Royal Society A: Mathematical, Physical and Engineering Sciences}} \textbf{\bibinfo{volume}{472}} (\bibinfo{year}{2016}).

\end{thebibliography}
